\documentclass[aps,showpacs,preprintnumbers,amsmath,amssymb,10pt,twocolumn,oneside]{article}
\usepackage{graphicx,subfigure}
\usepackage{ifpdf}
\usepackage[sort&compress,super]{natbib} %achica las referencias
\usepackage{caption}
\usepackage{color}
\usepackage{float} % para usar [H]
\usepackage{multirow, array} % para las tablas
\usepackage{rotating}
\usepackage{epstopdf}
\newcommand{\bea}{\vspace{0.25cm}\begin{eqnarray}}
\newcommand{\be}{\begin{equation}}
\newcommand{\ee}{\end{equation}}
\newcommand{\eea}{\end{eqnarray}}

\def\PRL{{Phys. Rev. Lett.} }
\def\PRA{{Phys. Rev.} A }
\usepackage[nottoc,notlof,notlot]{tocbibind} 
\begin{document}

\title{Quantum and classical characterization of single/few photon detectors}
\author{M. G. Mingolla $^{1,2,3}$, F. Piacentini $^1$, A. Avella $^{1,4}$, M. Gramegna $^1$,\\  L. Lolli$^1$,  A. Meda $^1$, I. Ruo Berchera $^1$, E. Taralli$^1$, P. Traina $^1$\\ M. Rajteri$^1$, G. Brida $^1$, I. P. Degiovanni$^1$ and M. Genovese$^{1}$\\ 
\\
$^1$ INRIM, Strada delle cacce 91, I-10135 Turin, Italy.\\
$^2$ Dipartimento di Fisica, Politecnico di Torino, Corso Duca \\ Degli Abruzzi 24,  Torino 10129, Italy \\
$^3$ Instituto Nacional de Tecnologia Indutrial (INTI), \\ Av Gral Paz 5445, B1650KNA
 San Mart\`in, Argentina\\
 $^4$ Dipartimento di Fisica, Universit\`a degli Studi di Torino,\\ via P. Giuria 1, 10125 Torino, Italy}

\maketitle

\begin{abstract}
This paper's purpose is to review the results recently obtained in the Quantum Optics labs of the National Institute of Metrological Research (INRIM) in the field of single- and few-photon detectors calibration, from both the classical and quantum viewpoint.\\
In the first part of the paper is presented the calibration of a single-photon detector with absolute methods, while in the second part we focus on photon-number-resolving detectors, discussing both the classical and quantum characterization of such devices.
\end{abstract}

%\tableofcontents
\section{Introduction}

% % % % % % % % % % % % % % % % % % % % % % % % % % % % % % % % % % % % % % % % % % % % % % % % % % % %

In the last decades, quantum optics experiments based on intensity light measurements have been realized mainly with intense (macroscopic) fields or at single-photon level, while photon counting with few-photon light (up to 100 photons) is a rather unexplored measurement regime. In the first two regimes (i.e. the macroscopic one and the single-photon level), several quantum optical applications have been derived, such as the ones related to quantum mechanics foundations investigation \cite{Genovese2005,Croca2013} quantum communication \cite{QMAVELLA,Scarani2009,Gisin2002,Braunstein2005}, computation \cite{Ladd2010,Kok2007} and metrology \cite{Zwinkels10,Giovannetti2011}. This partially derives from the fact that the output of detectors operating in this regime are not considered trustworthy, also because this region has not yet been investigated from the metrological point of view in order to provide well established characterization techniques.

In this paper we review the Klyshko calibration method, based on the parametric down conversion (PDC) phenomenon, for single-photon detectors, as well as few extensions to photon-number-resolving detectors (PNRD) realized in our laboratories; further works can be found in \cite{VV1,VV2}. On the other hand, applications of PDC light to analog regimes \cite{josab,JMO2009,optex2008,Lindenthal,Agafonov} are beyond the purpose of this review.

In the following, the term ``single-photon detector" refers to detectors producing just a ``click" irrespective of the number of photons impinging on it, e.g. Single-Photon Avalanche Diodes (SPADs) operating in Geiger mode.
On the contrary, typical detectors able to observe more than one photon are, e.g., high gain photomultiplier tubes \cite{burle,Allevi2010,zhang2012}, hybrid photodetectors  \cite{NIST}, CCDs and Electron Multiplying CCDs (EMCCDs) \cite{Haderka2005,Kalash2011}, Silicon Photomultipliers, the superconducting Transition Edge Sensors (TESs) \cite{lapo2011,fuk11,pre09,cab08,lit08,ros05,bandler06}, time-multiplexed single-photon detectors \cite{multiTemporal,fitch2003,rehacek2003,hederka2004} and single-photon detectors in tree configurations \cite{multiSpatial,multiSpatial2}.

No direct comparison upon common set of figures of merit was performed so far on these group of  PNRDs, also because at first it is necessary to establish a set of standardized definitions, connected with such figures of merit, as well as the corresponding characterization protocols.
In the first part of this article we will focus on the detection quantum efficiency, $\eta_{DUT}$ (where DUT stands for device under test), defined as the overall probability of observing the presence of a single photon impinging on the DUT. In the following we will see how to extend the Klyshko's absolute measurement technique (named Klyshko two-photon technique, KTPT) based on correlated photons obtained from PDC, originally developed for single-photon detectors \cite{Burnham1970,Klyshko1977,Kwiat1994,Migdall1995,Dauler1998,Brida2000,Castelletto2002,Ghazi2000,Migdall2002,Polyakov2007,jessica,penin1991,brigrn2000,gram2006,malygin,Kly1980,Bowman1986,Ginzburg1993,cheung2004},
to PNRDs. In order to characterize the detection behavior of PNRDs, together with the estimation of $\eta_{DUT}$ it is fundamental to provide a theoretical model for their measurement process.

Usually in quantum optics it is assumed that the detection model of a non-ideal detector (i.e. detector with $\eta_{DUT}\neq 1$) can be described as an ideal detector with unity efficiency placed after a beam splitter of transmissivity $\eta_{DUT}$. In practice, this implies that the detection process in the presence of more than one photon is described by the Bernoulli distribution.

This detection model is absolutely reasonable for typical ``analog" detectors for ``macroscopic" light. On the contrary, detection models for PNRDs may significantly differ from the Bernoulli one, as in the case of time-multiplexed single-photon detectors, trees of single photon detectors, or silicon photomultipliers.

In the following two PNRDs will be considered: the TES \cite{gram2006} and a tree of two single photon detectors. Since TES is essentially a superconductive microcalorimeter with a linear response, it is absolutely reasonable to expect for it a Bernoulli detection model. In the case of the tree of SPAD detectors, the model is no longer linear and will be analysed in the last section. 

In Section 2 we will present the KTPT applied to an Avalanche Photo Diode (APD) based single photon counting module, showing the detection efficiency obtained with this absolute calibration technique. 

Finally, in section 3 we will discuss the extension of the KTPT from the calibration of single-photon detectors to PNRDs. In particular, we will discuss a  KTPT-based calibration method, that exploits the whole information from the output of the PNRD, without referring to some specific detection model.

As already pointed out, PNRD characterization exploiting the above mentioned calibration techniques may not provide complete information regarding the detector behavior. On the contrary, at least in one case, the determination of the quantum efficiency is strongly dependent on the assumption of a specific detection model. For this reason, we consider also few calibration techniques providing the full characterization of the detection process of PNRDs. Here, the detection process is considered as a quantum operation, thus the technique consists in realizing the tomography of the quantum operation. 

\section{Two-photon Klyshko's calibration method}

The method using PDC to calibrate detectors is a well-established technique \cite{Burnham1970,Klyshko1977,Kwiat1994,Migdall1995,Dauler1998,Brida2000,Castelletto2002,Ghazi2000,Migdall2002,Polyakov2007,jessica,penin1991,brigrn2000,gram2006,malygin,Kly1980,Bowman1986,Ginzburg1993,cheung2004}. It has the peculiarity of being intrinsically absolute, as it does not rely on any externally calibrated radiometric standards \cite{Migdall1995}. The PDC phenomenon was predicted in 1961 by Louisell et al. \cite{Luisell1961}, and the very first experiment to observe coincidences between downconverted photons in 1970 \cite{Burnham1970} also included the first detector calibration using a PDC source. The method was not widely disseminated, however, and 7 years later Klyshko \cite{Klyshko1977} independently proposed that PDC could be used to measure detection efficiency. In the early 1980's, few groups \cite{Kwiat1994,Migdall1995,Dauler1998,Brida2000,Rarity1987,brigenovese2005} pushed the technique from demonstrational experiments to more accurate calibrations. As a consequence the KTPT has been added to the toolbox of primary radiometric techniques for detectors calibration \cite{Zwinkels10}, even if it has been deeply studied only in the case of single-photon detectors \cite{Castelletto2002,Ghazi2000,Migdall2002,Polyakov2007,jessica}.

\subsection{Theory}
The PDC process is used to create the correlated pair of photons allowing the absolute quantum efficiency measurement \cite{8,Burnham1970,10,11,gram2006,brideg2008}. The detection of one of the photons pairs announces the presence of its mate, and any missed detection (in absence of losses) of the announced photon is due to the non-ideal quantum efficiency of the DUT.

\begin{figure}
\centering
\includegraphics[width=8cm]{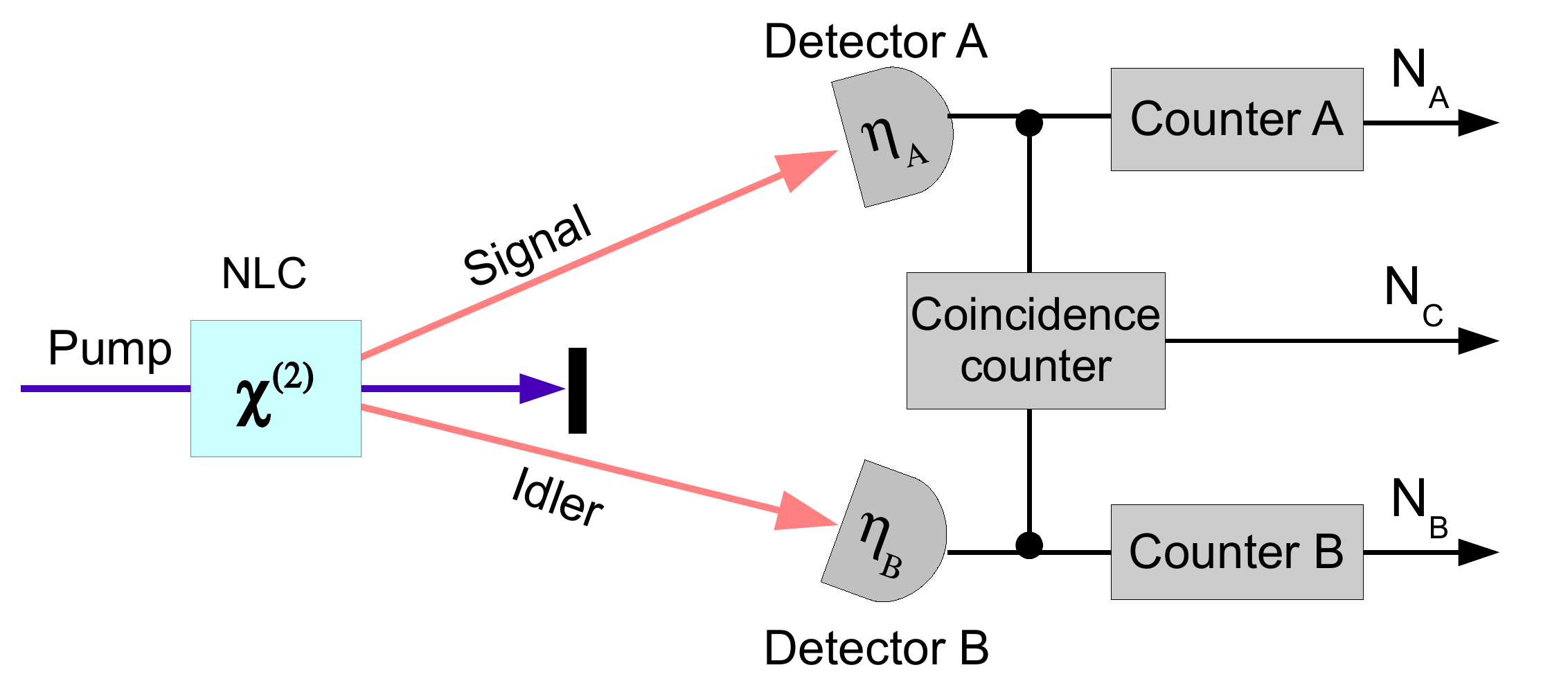}
\caption{Scheme for absolute calibration of a photon detector. A pump laser impinges on a nonlinear crystal generating PDC photons. Detector A with efficiency $\eta_A$ and detector B with efficiency $\eta_B$ collect the photons of correlated arms. Counters and coincidence electronics allow to obtain the number of signal counts $N_A$, the number of idler counts $N_B$ and the number of photons arriving in coincidence to both detectors $N_C$.}
\label{schemasetup}
\end{figure}

 The calibration scheme is depicted in Fig \ref{schemasetup}. Two correlated channels of PDC emission (dubbed signal and idler), are selected and directed to photo counters A and B respectively. In the ideal situation (no losses), the detection of one photon of the pair guarantees with certainty the presence of the second photon along the correlated direction. If $N$ is the total number of pairs emitted by the crystal in a given time interval, $N_A$ and $N_B$ are the average count rates recorded by detectors A and B during the same time interval, and $N_C$ is the coincidences count rate, we have the relations

\begin{equation}
\label{NaNb}
\begin{array}{c}
N_A=\eta_A(\lambda_A)N\\
N_B=\eta_B(\lambda_B)N\\
\end{array}
\end{equation}
where $\eta_A(\lambda_A)$ and $\eta_B(\lambda_B)$ are the detection efficiencies  of photodetectors A and B at specific wavelength $\lambda_A$ and $\lambda_B$. Due to the statistical independence of the detectors, the number of coincidences is

\begin{equation}
\label{coincidencias}
N_C=\eta_A(\lambda_A)\eta_B(\lambda_B)N
\end{equation}
then, the detection efficiency can be found as

\begin{equation}
\label{eficiencias}
\begin{array}{c}
\eta_A(\lambda_A)=\frac{N_C}{N_B}\\
\eta_B(\lambda_B)=\frac{N_C}{N_A}\\
\end{array}
\end{equation}

Anyway, in practice it is very difficult to guarantee that both detectors see only correlated photons, thus we have to associate each arm with a particular purpose: one detector is the device under test (DUT), while the other acts as a trigger, to indicate when a detection is expected in the DUT. We underline that, since the determination of $\eta_{DUT}$ is independent of the trigger efficiency, losses in the trigger channel do not affect the calibration technique

\subsection{Experimental setup}

In this section we analyze some details of the experimental realization of KTPT.

\begin{figure}
\centering
\includegraphics[width=8cm]{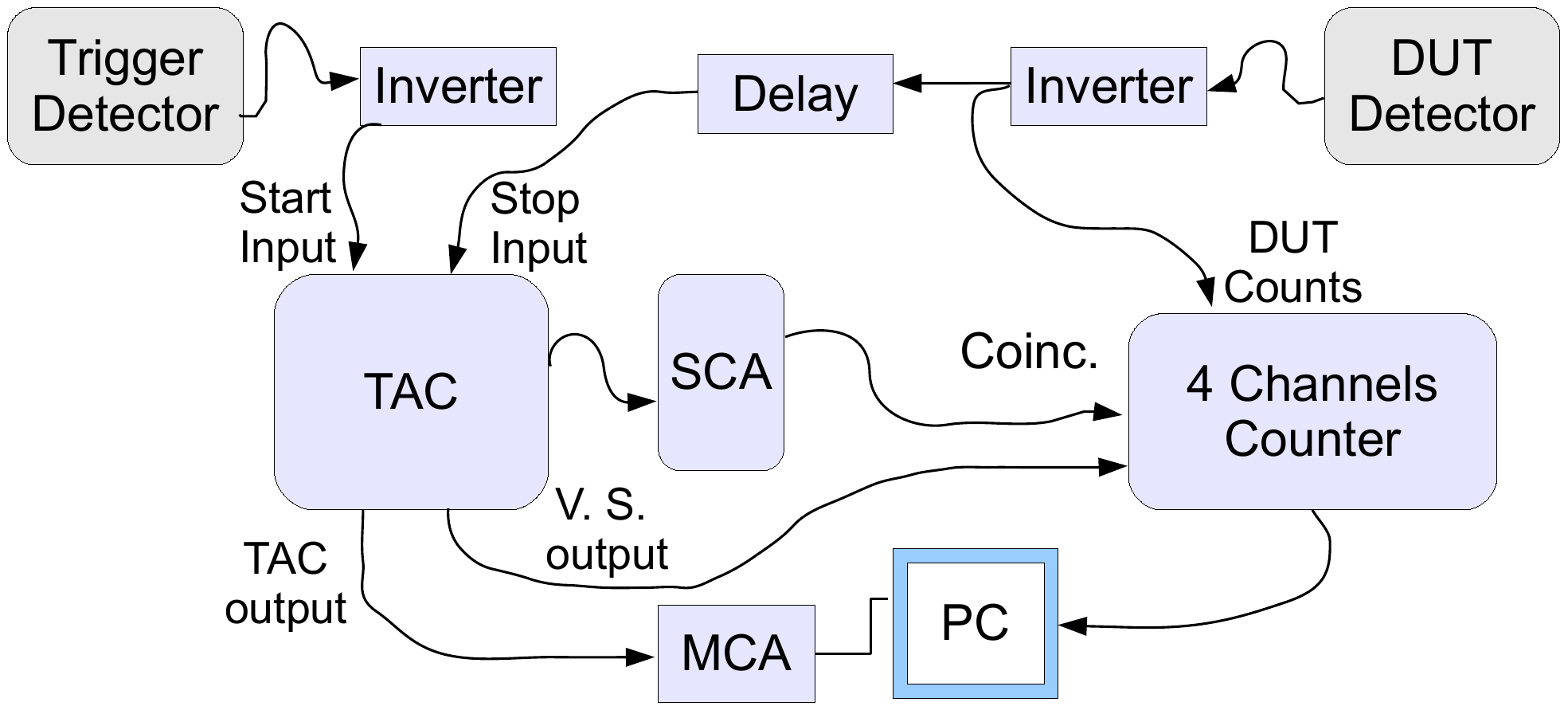}
\caption{Electronics setup. The output signal pulses of the detectors are inverted and properly delayed and sent to the TAC. The TAC outputs are sent to a MCA and to a SCA. TAC valid start (V.S.) output, the coincidence counts and DUT raw counts are measured by the counter. The whole measurement system is PC controlled.}
\label{electronica}
\end{figure}

Usually, the coincidence and counting electronics associated to these calibration experiments is like the one reported in Fig. \ref{electronica}.

The output signal from the trigger detector is sent to the start input of the TAC, and the DUT output is delayed (6.5 ns) and sent to the stop input of the TAC. The TAC output is sent simultaneously to a multichannel analyzer (MCA) and to a single-channel analyzer (SCA). The MCA records histograms of inter-arrival times of the DUT and trigger events. The SCA output is addressed to a counter in order to measure coincidence counts. Correlated photon pairs are seen in the histogram as a peak on top of a flat background resulting from uncorrelated output pulses from the two detectors. True coincidences are found by counting the events within a fixed time window around this peak and subtracting the flat background level within the same time window (referred to as accidental coincidences).

\subsection{Measurement procedure} \label{secc_measur}

To account for the presence of unwanted counts, Eq. \ref{eficiencias} has to be modified. In addition to the correlated photons, each detector suffers from background counts, due to unwanted external light (e.g. stray light or unheralded PDC light), and spurious counts due to thermal fluctuation inside the detector or trapped carriers (dark counts and afterpulses). Thus, spurious coincidence counts are superimposed on the correlated pairs, leading to the above mentioned background counts and accidental coincidences. To correct for the unwanted detected light, the measured quantum efficiency, $\eta_{DUT}^{meas}$, is estimated from \cite{10,Kwiat1994,Brida2000} 

\begin{equation}
\label{effDUTmeasu}
\begin{array}{c}
\eta_{DUT}^{meas}=\frac{\left\langle m_c\right\rangle - \left\langle A\right\rangle}{\left\langle m_{vs}\right\rangle - \left\langle m_{B}\right\rangle}\\
\end{array}
\end{equation}
where $\left\langle m_c\right\rangle$ are the average coincidence counts measured by TAC/SCA, $\left\langle m_{vs} \right\rangle$ the average valid start counts, $\left\langle m_{B}\right\rangle$ the average background counts on the valid start and $\left\langle A\right\rangle$ are the accidental coincidence counts. 

Concerning this last correction one has to detail a little more. The TAC valid start output provides only the true trigger counts that are considered for conversion and give contribution to coincidences. Thus the TAC  dead time effect can be neglected thanks to the valid start output. We should also note that the number of valid start counts able to produce an accidental coincidence drastically changes if the peak of coincidences is in the TAC windows or not. Because the accidental counts are evaluated by adding a delay to the DUT output, in order to move the peak out of the measurement window, a correction for $\left\langle A\right\rangle$ should be added accounting to this valid start mismatch. A reasonable first order correction is given by

\begin{equation}
\label{A}
\begin{array}{c}
\left\langle A'\right\rangle\cong\left\langle A\right\rangle \frac{\left\langle m_{vs}^{in}\right\rangle}{\left\langle m_{vs}^{out}\right\rangle}
\end{array}
\end{equation}
where $\left\langle m_{vs}^{in}\right\rangle$ is the average value of the valid start counts when the coincidence peak is in the TAC window and $\left\langle m_{vs}^{out}\right\rangle$ is the valid start average when the coincidence peak is not in the TAC windows.

Then, the measured quantum efficiency becomes 

\begin{equation}
\label{B}
\eta_{DUT}^{meas'}=\frac{\left\langle m_c\right\rangle - \left\langle A\right\rangle\frac{\left\langle m_{vs}^{in}\right\rangle}{\left\langle m_{vs}^{out}\right\rangle}}{\left\langle m_{vs}^{in}\right\rangle - \left\langle m_{B}\right\rangle}
\end{equation}

If we take into account a correction due to optical losses, we obtain the quantum efficiency of just the detector under calibration

\begin{equation}
\label{effDUT2}
\begin{array}{c}
\eta_{DUT}=\frac{1}{\tau_{DUT}}\eta_{DUT}^{meas'}\\
\end{array}
\end{equation}
where $\tau_{DUT}$ is the total transmittance of the DUT arm. A careful estimation of $\tau_{DUT}$ is crucial in order to adequately evaluate the quantum efficiency of the device under calibration.

\subsection{Experimental Results}
This kind of protocol has been experimentally implemented many times (e.g. Ref,\cite{Ghazi2000}).

According to Ref.s [\cite{10,Kwiat1994}], the statistical uncertainty associated with this two-photon measurement technique is deduced by applying the uncertainty propagation law  \cite{16} to the model of Eq. \ref{effDUT2}

\begin{equation}
\label{effDUTmeas_incertidumbre}
\begin{array}{c}
u^2(\eta_{DUT})=c_1^2 u^2(m_c)+c_2^2 u^2(A)+c_3^2 u^2(m_{vs}^{in})+\\
c_4^2 u^2(m_{vs}^{out})+c_5^2 u^2(m_B)+c_6^2u^2(\tau_{DUT})+\\
2 \rho_{1,3} c_1 c_3  \sqrt{u^2(m_c)u^2(m_{vs}^{in})}+\\
2 \rho_{2,4} c_2 c_4  \sqrt{u^2(A)u^2(m_{vs}^{out})}
\end{array}
\end{equation}
where $u^2(x)=\left\langle x^2\right\rangle-\left\langle x \right\rangle^2$ is the variance of a generic variable $x$. Sensitivity coefficients $c_i$ are deduced by standard uncertainty propagation rules and the correlation coefficients $\rho_{i,j}$ are evaluated from repeated experimental data as $\rho_{i,j}=(\left\langle x_i x_j \right\rangle - \left\langle x_i \right\rangle\left\langle x_j \right\rangle)/\sqrt{u^2(x_i)u^2(x_j)}$.

Recently, a new implementation of the KTPT protocol has been carried out, obtaining a value of $\eta_{DUT}$ with a relative uncertainty of $10^{-4}$ (Ref. \cite{mingolla}), paving the way to applications in photometry and metrology \cite{Brida2000,Ghazi2000,10,castelletto2000}.

\section{Extension of KTPT to PNRDs}

The extension of the KTPT to PNRDs system is quite straightforward \cite{Avella2011}. 

\begin{figure}[h]
\begin{center}
 \includegraphics[width=8cm]{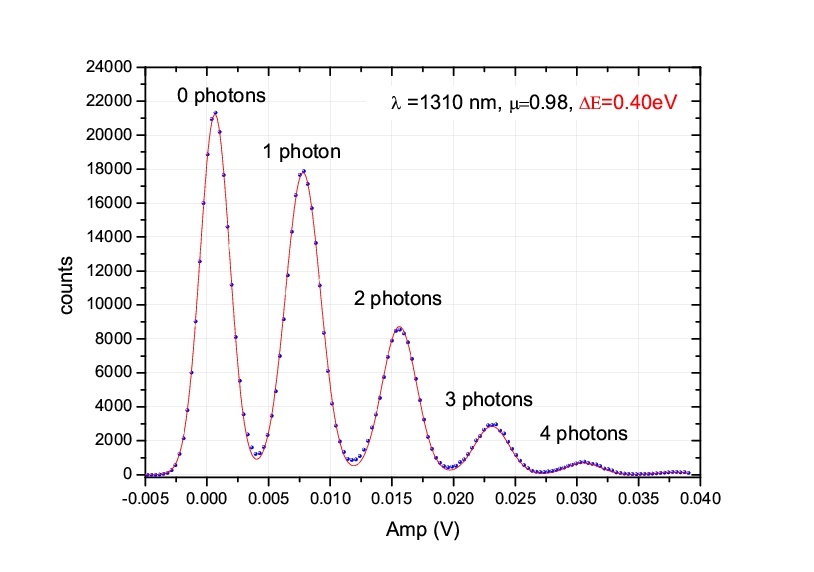}
\end{center} \caption{ The figure shows a TES measurement of Poissonian statistics with a mean photon number per pulse of 0.98 at 1310 nm. The line shows a plot of best-fit to the data, convolving the Poissonian distribution with the energy resolution of the TES \label{peaks}.}
\end{figure}

As in the classic KTPT, it is necessary to perform two separate measurements in the presence and in the absence of the heralded photon. For each measurement, a data histogram analogous to the one reported in Fig. \ref{peaks} is obtained, from which it is possible to estimate the probabilities of observing $i$ photons per heralding count in the presence of a heralded photon ($P(i)$) and in the absence of the heralded photon ($\mathcal{P}(i)$).
The probability of observing 0 photons in the presence of the heralded photons is simply the product of the probability of non-detection of the heralded photons multiplied by the probability of having 0 accidental counts:
 
\begin{equation} 
P(0)= (1-\eta) \mathcal{P}(0),
\label{p0}
\end{equation}
where $\eta$ is the quantum efficiency of the DUT channel, including the optical losses, i.e.  $\eta= \tau_{DUT} \eta_{DUT}$.

Analogously, the probability of observing $i$ counts in presence of a heralded photon can be written as
\begin{equation}
\begin{array}{c}
  P(i)= (1-\eta) \mathcal{P}(i)+\eta \mathcal{P}(i-1) \\
  \mathrm{with} ~ i=1,2,...
\end{array}
 \label{pi}
\end{equation}

From these equations, we can derive $i+1$ ways to evaluate the quantum efficiency of our PNRD: from Eq.(\ref{p0}) the efficiency can be estimated as
\begin{equation} \label{p02}
\eta_0= \frac{\mathcal{P}(0)-P(0)}{\mathcal{P}(0) },
\end{equation}
while from Eq. (\ref{pi})
\begin{equation} \label{pi2}
\eta_i= \frac{P(i)-\mathcal{P}(i)}{\mathcal{P}(i-1)-\mathcal{P}(i) }.
\end{equation}

It is noteworthy to observe that the set of hypotheses in the context of this calibration technique is exactly the same as the ones of the proper KTPT, and the same measurements are necessary. In this case, for each value of $i$ (i.e. for each peak of Fig. \ref{peaks}) we obtain an estimation for $\eta$, allowing also a test of consistency for the estimation model. Furthermore, as the number of heralding counts appears both at the numerator and at the denominator of Eq.s (\ref{p02}) and (\ref{pi2}), the correction for false heralding counts coming from stray-light, dark counts and afterpulses can be neglected.

\subsection*{TES calibration}

In ref. \cite{Avella2011} is shown a recent implementation of the proposed calibration technique applied to a TES. For this purpose, a certain number of the detector properties should be considered, e.g.:

\textit{TES jitter}: the poor temporal resolution of TESs (time jitter larger than 100 ns) does not allow the use of small coincidence temporal windows. A reasonable solution is the exploitation of a heralded single photon source and the measurement of the coincidences in presence and in absence of heralding signal. The second measurement allows a trustworthy estimation of the accidental coincidences, since TESs are not affected by afterpulsing.

\textit{TES deadtime}: as a single-photon detector, TES behaves like a paralizable detector with extending dead-time \cite{castelletto2000}, such as a photomultiplier operating in Geiger mode. To avoid deadtime distortion in the statistics of measured counts, it is necessary to use a pulsed heralded single-photon source with period larger than the deadtime of the detector, in order to avoid unwanted photons impinging on the TES surface during the deadtime, whose only effect would be the deadtime extension.

\textit{Optical losses}: in the KTPT a careful estimation of the optical losses on the arm of the detector under test is crucial. In the specific case of the TES, where the detector input is the pit of the optical fiber, optical losses should account also for the coupling efficiency in the optical fiber.

\medskip

\begin{figure}[h]
\begin{center}
 \includegraphics[width=8cm]{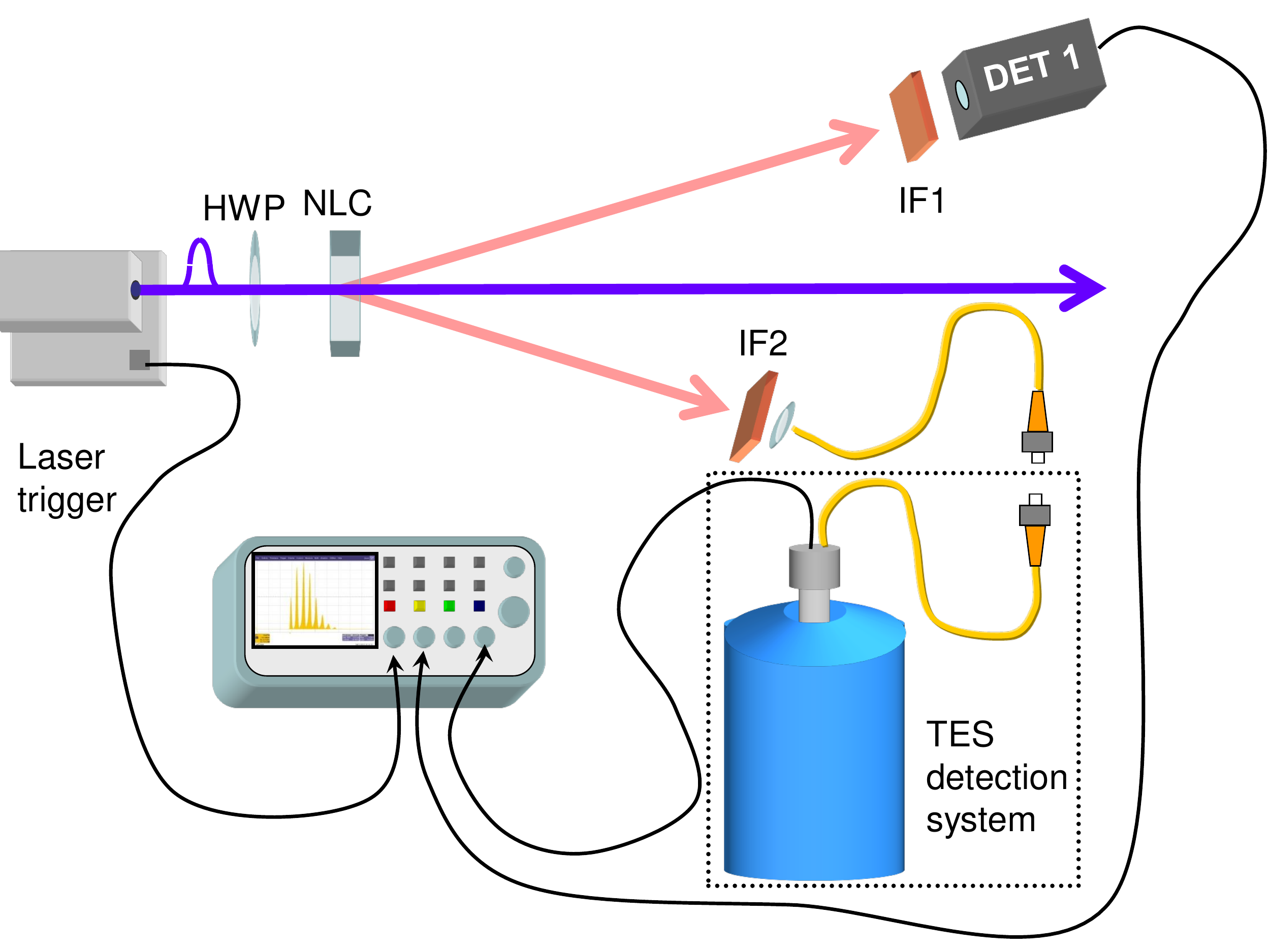}
\end{center} \caption{Experimental setup for the measurement efficiency of a TES detector. A non-linear crystal pumped by a pulsed laser at 406 nm generates non-collinear degenerates PDC, that is used as a heralded single photon sources. The presence of a photon in the trigger detector (DET1) announces the presence of a photon in the conjugated direction, that is coupled in a single mode optical fiber and sent to the TES.
\label{TESsetup}.}
\end{figure}

Our TES sensor consists of a superconducting Ti film proximised by an Au layer \cite{fab}. This detector has been thermally and electrically characterised by impedance measurements \cite{taralli}. The transition temperature of the TES, voltage biased \cite{etf} and mounted inside a dilution refrigerator at a bath temperature of 40 mK, is $T_c$=121 mK.
In Fig. \ref{TESsetup} the experimental setup used in the TES calibration is shown. A single-mode optical fiber illuminates the TES active area (20 $\mu$m x 20 $\mu$m). A stereomicroscope is used to align the fiber on the TES \cite{lapo}, being the distance between the fiber tip and the detector approximately 150 $\mu$m. The read-out is based on a DC-SQUID array \cite{squid} coupled to a digital oscilloscope for signal analysis. The energy resolution is $\Delta E_{FWHM}= $ 0.4 eV, with a response time of 10.4 $\mu$s \cite{lapo}.

A heralded single photon source is used to perform the calibration: a type-I BBO crystal is pumped by a pulsed laser at 406 nm  generating degenerate non-collinear PDC \cite{11,gram2006,qurad2}. A heralding photon at 812 nm is detected by a single photon detector (DET1 in fig.\ref{TESsetup}) announcing the presence of the conjugated photon (812 nm) in the conjugated direction. The announced photon is sent to the TES detector by means of a single-mode optical fiber. Because low repetition rate is needed to avoid pile up effect in the statistics of measured counts, the pump laser is electrically driven by a train of $80$ ns pulses with a repetition rate of 40 kHz. 

It is easy to estimate the probabilities $P(i)$ and $\mathcal{P}(i)$, measuring the events seen by the TES in the presence or absence of heralded signal. Defining $C(i)$ and $\mathcal{C}(i)$ as the $i$ photons events observed by the TES in the presence or absence of the heralding photon respectively, \mbox{$P(i)= C(i)/\sum_{i} C(i)$} and \mbox{$\mathcal{P}(i)=\mathcal{C}(i)/\sum_{i} \mathcal{C}(i)$}.

\begin{figure}[tbp]
\par
\begin{center}
\includegraphics[angle=0, width=8cm]{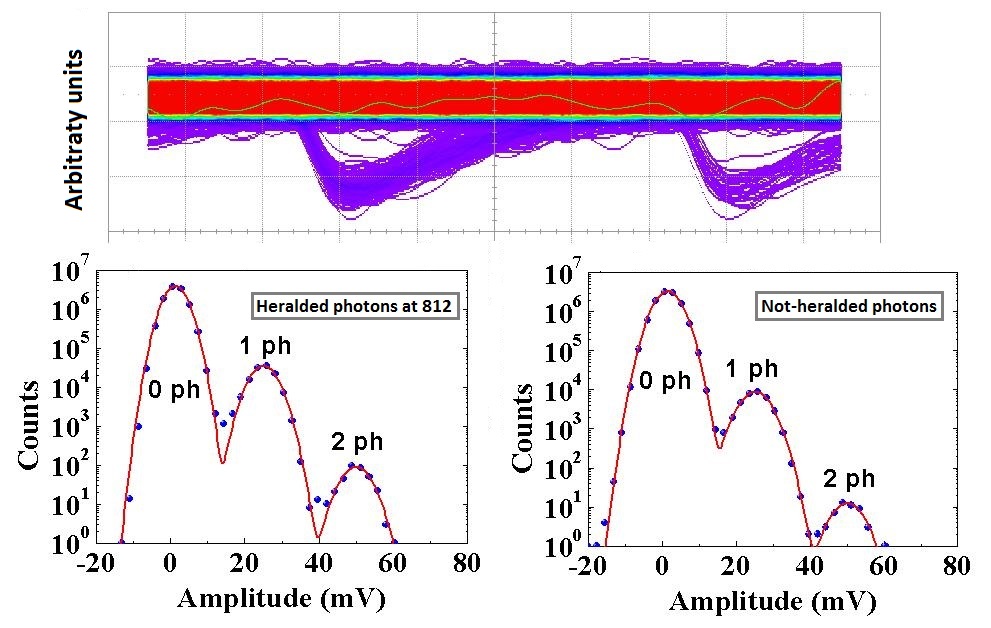}
\caption{Experimental data: oscilloscope screen-shot with traces of the TES detected events. The group of traces on the left (right) are obtained in the presence (absence) of heralding signals.
Insets (a) and (b) present the histogram of the amplitudes of the pulses in the presence and in the absence of heralding photons, together with their gaussian fits.} \label{data}
\end{center}
\end{figure}

TES events are detected by an oscilloscope, typical traces are reported in fig. \ref{data}. The oscilloscope readout is triggered only when both the pump laser trigger and the heralding detector clicks are present. In order to measure both the events in the presence and in absence of the heralded photon, the time base is set to show two consequent laser pulses (corresponding to the  left and right pulses on fig. \ref{data}, respectively).
A histogram is generated measuring the amplitude of the pulses recorded on the oscilloscope, each peak corresponds to a different number of detected photons. As an example, the right graph in fig. \ref{data} shows the histogram of a heralded event while in the left graph the histogram of an event in the absence of heralded photon is reported. To estimate parameters $C (i)$ and $\mathcal{C}(i)$, the histograms are fitted with Gaussian curves and the integral of each peak is calculated.

In a more realistic model, we can re-write equations \ref{p0} and \ref{pi} considering the presence of false heralding photons due to dark counts and stray light arriving to the trigger detector. If we denote by $\xi$ the probability of having a true heralding count then, the probability in eq. \ref{p0} of observing no photons on the PNR detector is

\begin{equation} \label{p0bis}
\begin{array}{c}
P(0)= \xi [(1-\eta) \mathcal{P}(0)]+ (1-\xi) \mathcal{P}(0) \\
=(1-\xi  \eta) \mathcal{P}(0)
\end{array}
\end{equation}
and the efficiency 

\begin{equation} \label{p02bis}
\eta_0= \frac{\mathcal{P}(0)-P(0)}{\xi \mathcal{P}(0)}
\end{equation}
while the probability of observing $i$ counts is
\begin{equation} \label{pibis}
P(i)= \xi[(1-  \eta) \mathcal{P}(i)+ \eta \mathcal{P}(i-1)]+ (1-\xi)\mathcal{P}(i) 
\end{equation}
and the efficiency can be estimated as
\begin{equation} \label{pi2bis}
\eta_i= \frac{P(i)-\mathcal{P}(i)}{\xi(\mathcal{P}(i-1)-\mathcal{P}(i))}
\end{equation}

Six repeated measurement of five hours, corresponding approximately to $11\times 10^6$ heralding counts were made. By measuring both the number of events triggered by the laser pulses and detected by DET1 in the presence of PDC emission, $n_p$, and in the absence of PDC light, $n_a$, the probability of having true heralding counts is calculated as $\xi=(n_p - n_a)/n_p$, obtaining $\xi=0.98793\pm 0.00007$.

The histograms of fig. \ref{data} show that only the peaks corresponding to the detection of zero, one and two photons have enough counts to be identified, while peaks corresponding to the arrival of three or more photons simultaneously to the TES are negligible. Thus, three different values of the quantum efficiency were estimated: 
 $\eta_{0}= (0.709 \pm 0.003) \%$, $ \eta_{1}= (0.709 \pm 0.003) \%$, and $\eta_{2}= (0.65 \pm 0.05) \%$.
An exhaustive uncertainty analysis is given in table \ref{tableTES} \cite{16}. The three values obtained for $\eta$ are consistent with each other, since the TES has been recently proved to be a linear detector \cite{tomo}, as generally believed \cite{had,had1}.
In average, the result for the efficiency of the TES channel is $\eta= (0.709\pm 0.002)\%$.

\begin{table}

 \caption{Uncertainty contributions of the different quantities (Q) to the measurement of $ \eta_{0}$, $ \eta_{1}$ and $ \eta_{2}$. The uncertainties are calculated according to the Guide of Uncertainty Measurement \cite{16} with $k = 1$.} \label{tableTES}
\begin{center}
\scalebox{0.7}[0.8]{
\begin{tabular}{|c|c|c|c|c|c|}
\hline

Q& Value& Stand.  & \multicolumn{3}{|c|}{ Unc. contribution (\%)}\\
\cline{4-6}
&  & Unc.&  $\eta_{0}$&$ \eta_{1}$&$ \eta_{2}$\\

\hline
$C_0$ & $5.069\times 10^6$ & $1.4\times 10^4$ & $-0.003$ & $-0.003$ & $-0.003$\\
$C_1$ & $5.0200\times 10^4$ & $200$ & $0.004$ & $0.004$ & $-4\times 10^{-5}$\\
$C_2$ & $118$ & $6$ & $2\times 10^{-4}$ & $-2\times 10^{-6}$ & $0.05$\\
$\mathcal{C}_0$ & $5.103\times 10^6$ & $1.4\times  10^4$ & $8\times 10^{-4}$ & $8\times 10^{-4}$ & $0.003$\\
$\mathcal{C}_1$ & $1.4600\times  10^4$ & $150$ & $-0.003$ & $-0.003$ & $-0.007$\\
$\mathcal{C}_2$ & $23.9$ & $1.5$ & $-3\times 10^{-5}$ & $3\times 10^{-7}$ & $-0.02$\\
$\xi$ & $0.98794$ & $7\times 10^{-5}$ & $-6\times 10^{-5}$ & $-6\times 10^{-5}$ & $-5\times 10^{-5}$\\
\hline
$\eta_{0}$ & $0.709\%$ & - & $0.003\%$ &-&-\\
$\eta_{1}$ & $0.709\%$ & - &-& $0.003\%$ &-\\
$\eta_{2}$ & $0.65\%$&  -&-&-& $0.05\%$ \\
\hline
\end{tabular}}
\end{center}
\end{table}

To provide a precise estimation of the bare TES quantum efficiency $\eta_{TES}$  a careful estimation of the optical transmittance $\tau$ is needed, accounting for the coupling efficiency in the optical fiber and the optical losses in the non-linear crystal. According to the results of Ref.s \cite{Polyakov2007,jessica}, one could provide an estimation of this parameter with a relative uncertainty better than 1\%.

The parameter $\tau$ is estimated to be $10 \%$. Concerning the TES sensor, on the basis of the material used the espected efficiency should be around 49\%. Geometrical and optical losses in the conection between the superconducting film and the outside of the refrigerator contribute to lower the value of $\eta_{TES}$ down to 7\%.

\section{POVM reconstruction technique for a TES detector}
The most general measurement in quantum mechanics is described by positive operator-valued measurement (POVM). The most general description of a PNRD is therefore provided by its POVM. A complete description of TES detectors is crucial for several applications  \cite{lss99,jar01,dem03,mit03,Zam05,zzam05,lob08,rah11,D'Ariano2004}. 

As stated before,  TESs are intrinsically phase-insensitive linear PNRDs, with a detection process corresponding to a binomial convolution and with not dark counts. 

Thus, we can assume that the elements of their POVM $\{\Pi_n\}$ are diagonal operators in the Fock basis, with completeness relation $\sum_n \Pi_n = \mathbf{I}$:
\begin{equation}
\Pi_n=\sum_m \Pi_{nm} |m\rangle \langle m|,
\end{equation}
The probability of detecting $n$ photons with $m$ photons impinging on the TES is described by the matrix elements 
 $\Pi_{nm}=\langle m| \Pi_n| m\rangle$.
By exploiting a technique \cite{h1,lun09,h2} based on recording the detector response for a known and suitably chosen set of input states, the characterization of the detector at the quantum level can be achieved. In order to carry out the tomography of the TES POVM ( i.e. to reconstruct the $\Pi_{nm}$), we use an ensemble of coherent probes providing a sample of the Husimi Q-function of the POVM elements.
\par
If we consider a set of $S$ coherent states of different amplitudes $|\alpha_j\rangle$, $j=1,...,S$, the probability of detecting $n$ photons, with the $j$-th state as input is given by 
\begin{equation}
\label{eq:stat_model}
p_{nj} =\hbox{Tr}[|\alpha_j\rangle\langle\alpha_j| \Pi_n] =\sum_m \Pi_{nm}\, q_{mj}
\end{equation}
where $q_{mj}=\exp(-\mu_j)\mu_j^m/m!$ is the photon statistics of the coherent state
$|\alpha_j\rangle$, $\mu _j = |\alpha_j|^2$ being its average number of photons.
By sampling the probabilities $p_{nj}$ and inverting the statistical model composed by the set of Eqs. (\ref{eq:stat_model}), the matrix elements $\Pi_{nm}$ are reconstructed. 

A sensible truncation on the Hilbert space of the $\Pi_{nm}$ should be chosen, e.g. with the constraint that, with the chosen set of coherent states, we have no significant data for $m \geq M$, and so we cannot investigate the performances of the detector in that regime.

\par
To solve the statistical model in (\ref{eq:stat_model}) maximum likelihood (ML) methods may be used. 
In our case, we have estimated $\Pi_{nm}$ by regularized minimization of the square difference
\begin{equation}
\sum_{nj} (\sum_{m=0}^{M-1} q_{mj}\, \Pi_{nm} - p_{nj})^2
\end{equation}
the physical constraints of ``smoothness'' are satisfied by exploiting a convex, quadratic and device-independent function \cite{lun09}. Normalization is forced ($\sum_{n=0}^{11} \Pi_{nm}=1,\;\forall m$), with the last POVM element defined as $\Pi_{11}=1-\sum_{n=0}^{10}\Pi_{n})$.
\par

\subsection{Experiment}
The TES sensor characterized in this experiment is composed by a $\sim$ \mbox {90 nm} thick Ti/Au film \cite{por08,tar07}, with an effective sensitive area of ($20\times 20$) $\mu$m.
The characterization has been performed in a dilution refrigerator with a base temperature of 30 mK.  
A DC-SQUID current sensor \cite{dru07} is used to read out operations, associated to room temperature SQUID electronics, addressed to an oscilloscope for the data acquisition.
TES is illuminated with a  fiber-coupled power-stabilized pulsed-laser at  $\lambda=1570$ nm (with a pulse duration of 37 ns and a repetition rate of 9 kHz).
The laser pulse is also used to trigger the data acquisition for a temporal window of 100 ns.
By using a calibrated power meter the laser pulse energy is measured ($(365\pm 2)$ pJ), and then attenuated to a range going from 130 to 6.5 photons per pulse in average, obtaining 20 different coherent states $|\alpha_j \rangle =|\sqrt{\tau_j} \alpha\rangle$ where $\tau_j$ is the
channel transmissivity, $j=1, ..., 20$. %%

\begin{figure}
     \begin{center}
        \subfigure[]{%
            \label{fig:first}
            \includegraphics[width=0.4\textwidth]{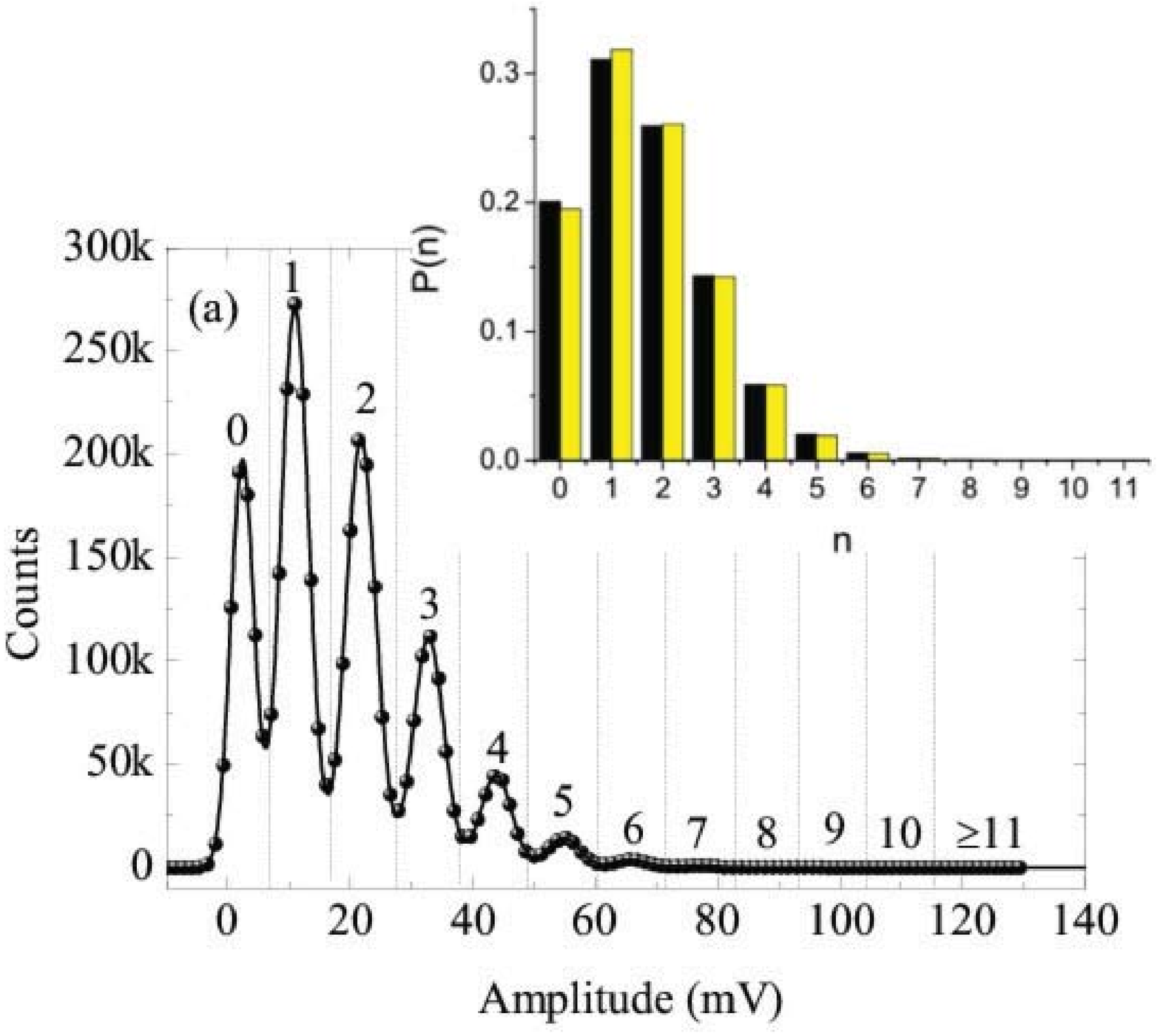}
        }\\
        \subfigure[]{%
            \label{fig:third}
            \includegraphics[width=0.4\textwidth]{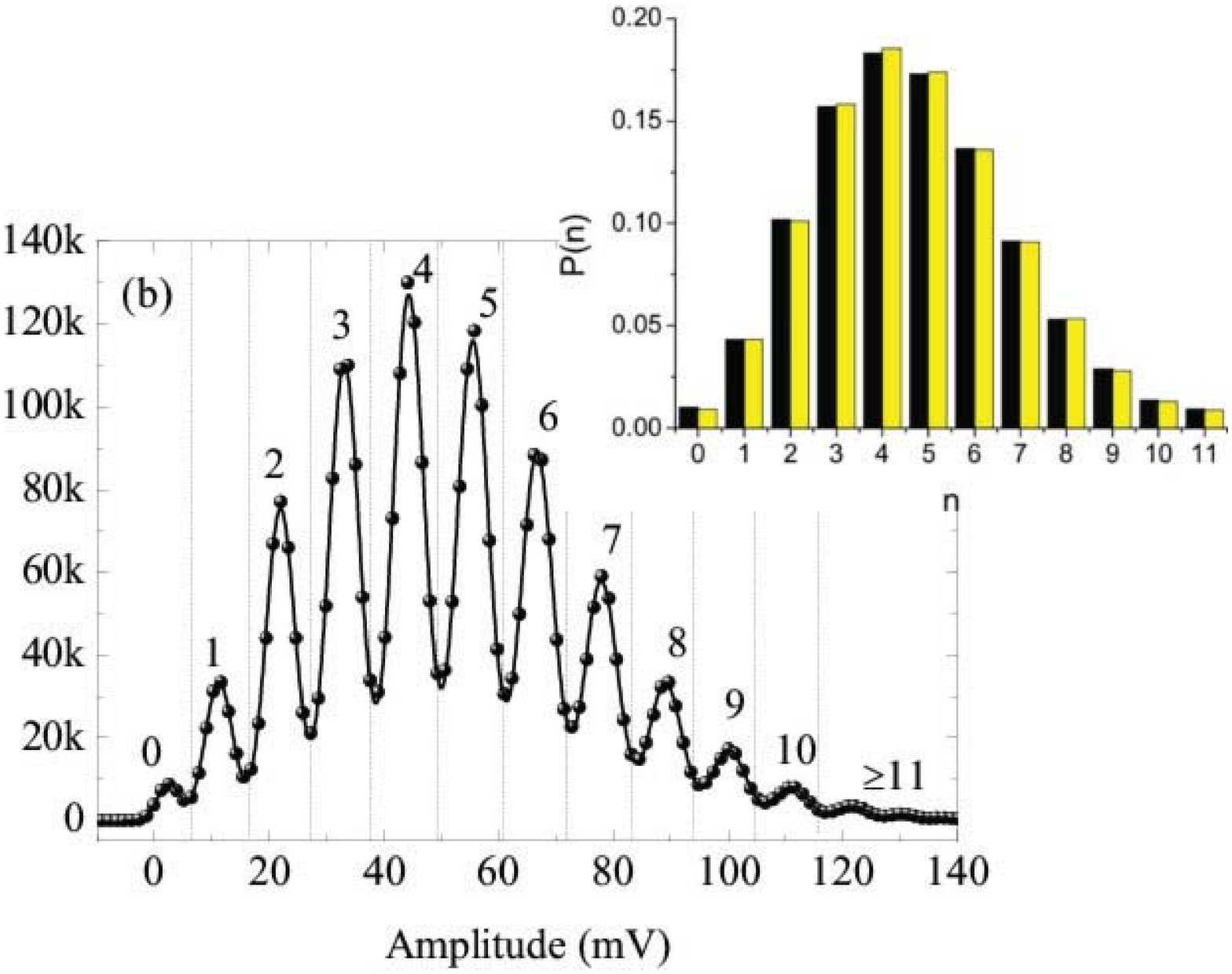}
        }    
    \end{center}
    \caption{Experimental data obtained with a coherent state characterized by mean photon number per pulse $\mu=31$ is shown in figure (a), while in figure (b) a state with $\mu=87$ is presented. Dots represent the TES counts, each point corresponds to a binning of an amplitude interval of $1.3$ mV. Solid lines are the Gaussian fits on the experimental data, while the dotted vertical lines are the chosen thresholds. The insets of both figures show a comparison of the experimental probability distribution (black bars), obtained from measurements binned according to the drawn thresholds, with the corresponding Poisson distributions of mean value $ \eta \mu$ (with $\eta= 5.1 \%$) (yellow bars): the experimental results are in remarkable agreement with the theoretical predictions, showing respectively a fidelity of
        $99.994\%$ and $99.997\%$.}
   \label{f1_TES}
\end{figure}

Since our source emits almost monochromatic photons, in ideal conditions a discrete energy distribution with outcomes separated by a minimum energy gap $\Delta E=\frac{hc}{\lambda}$ is expected.
Experimentally, a distribution with several peaks is observed, whose FWHM is determined by the energetic resolution of the TES.
 
In a first calibration run, we fit the data with a sum of independent Gaussian functions (Fig. \ref{f1_TES}); these fits allowed us to fix the amplitude thresholds (located close to the local minima of the fit) corresponding each to a different number of detected photons. The histogram of counts is obtained just by binning on the intervals identified by these thresholds. The distributions $p_{nj}$ are finally evaluated upon normalizing the histogram bars to the total number of events collected for each j.

\par
\subsection{Results}
\begin{figure}[htb]
\begin{center}
\includegraphics[width=1\columnwidth]{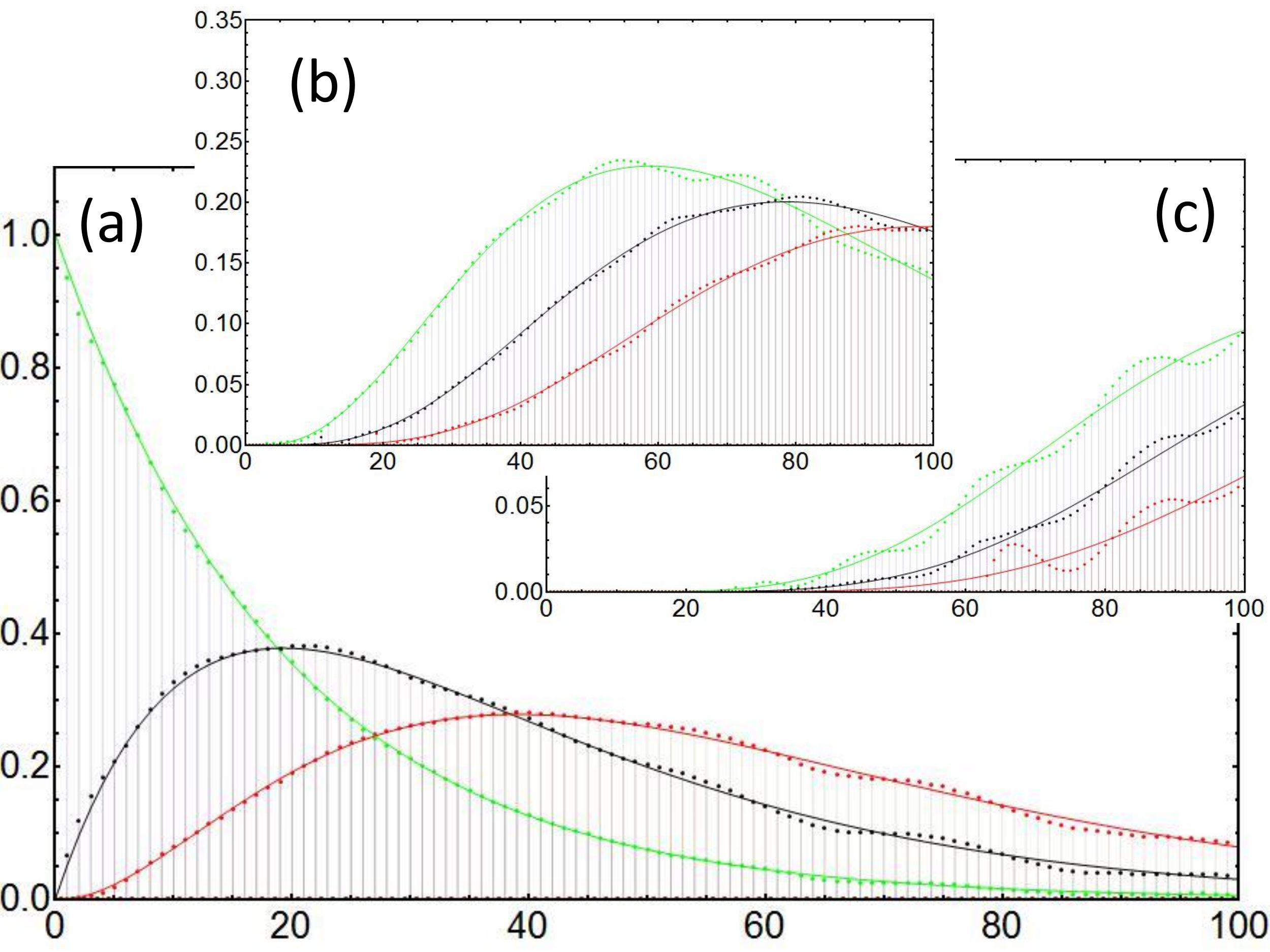}
\caption{Reconstructed POVM of our TES. Dots represent the matrix elements $\Pi_{nm}$ as a function of $m=0,...,100$ for $n=0,1,2$ (respectively green, black and red graphs in plot (a)), $n=3,4,5$ (b), $n=6,7,8$ (c). Continuous lines show the POVM of a linear photon counter with quantum efficiency $\eta=5.1\%$.} \label{f2_POVM}
\end{center}
\end{figure}

The POVM reconstruction of the TES detection system has been performed \cite{POVMTES} up to $M=140$ incoming photons and
considering $N=12$ POVM elements $\Pi_n$. The probability operator of more than 10
photons is given by $\Pi_{11}= \mathtt{1}-\sum_{n=0}^{10}\Pi_{n} $.
The matrix elements $\Pi_{nm}$ of the first 9 POVM operators ($n=0,..,8$),for $0\leq m\leq 100$ are shown in Fig. \ref{f2_POVM}.
As mentioned before, the POVM of a linear photon counter can be expressed as $\Pi_n=\sum_{m=n}^{\infty} B_{nm} |m\rangle\langle m|$  spectral measure with $B_{nm} =\left(\begin{array}{c} m \\ n \end{array}\right) \eta^n (1-\eta)^{m-n} $, where $\eta$ is the quantum efficiency of the detector. 
In order to compare the POVM elements of the linear detector ($B_{nm}$), with the reconstructed POVM elements ($\Pi_{nm}$) it is necessary to estimate first the value of the quantum efficiency $\eta$.

This can be done by averaging the values of $\eta$ which maximize the log-likelihood functions
\begin{equation}
L_j=\sum_{n} N_{nj} \log\left(\sum_m B_{nm} q_{mj}\right)
\end{equation}
where $N_{nj}$ is the number of $n$-count events obtained with the $j$-th input state $|\sqrt{\tau_j}\alpha\rangle$. 

Using this procedure, the estimated value of the quantum efficiency is $\eta=(5.10 \pm 0.04) \%$, where the uncertainty accounts for the statistical fluctuations. %%

\par
%%%
An excellent agreement between the reconstructed POVM and the linear one with the estimated quantum efficiency is observed in Fig. \ref{f2_POVM}. 

In particular, the elements of the POVM are reliably reconstructed for $m \leq 100$, whereas for higher values of $m$ the quality of the reconstructions degrades. In this regime ($m \leq 100$) the fidelity $F_m=\sum_n \sqrt{\Pi_{n m} B_{n m}}$ is larger than 0.99, while it
degrades to 0.95 for $100 \leq m \leq 140$.

Experimental uncertainties effects are investigated by performing a sensitivity analysis and taking into account the uncertainties on the energy of the input state and on the attenuators, obtaining fidelities always greater than $98.35\%$ for all the entries. In order to further confirm the linearity hypothesis, we have compared the measured distributions 
$p_{nj}$ with the ones computed for a linear detector, i.e.
\begin{equation}
l_{nj}= \eta^n \exp(-\eta \mu_j) \mu_j^n/n!
\end{equation}
 and with those obtained using the reconstructed POVM elements, i.e.
\begin{equation}
r_{nj}=\sum_{m=n}^{M} \Pi_{nm} q_{mj} .
\end{equation}
The excellent agreement between these three distributions (fidelities always above 99.5\%) confirms the linear behaviour of the detector, proving that the reconstructed POVM provides a reliable description of its detection process. %%
\par
Finally, to take into account the possible presence of dark counts, the detection model has been modified.
Assuming a Poissonian background, the matrix elements of the POVM are now given by $\Pi_{nm}= \exp(-\gamma) \sum_j\gamma^j/j!\,B_{(n-j)m}$; a ML
procedure has been developed to estimate both the quantum efficiency $\eta$ and the mean number of dark counts per pulse $\gamma$. With this procedure, we found that the value for $\eta$ is statistically indistiguishable from the one obtained with the linear-detector model, whereas the estimated dark counts per pulse are zero
within the statistical uncertainties, $\gamma=(-0.03 \pm 0.04)$, in excellent agreement with
the direct measurement performed on our TES detector ($\gamma=(1.4\pm0.6)\times10^{-6}$).

% % % % % % % % % % % % % % % % % % % % % % % % % % % % % % % % % % % % % % % % % % % % % % % % % % % % % % % % % % % % % % % % % % % % % % % % % % % % % % % % % % % % % % % % % % % % % % % % % % % % % % % % % % % % % % % % % % % % % % % % % % % % % % % % % % % % % % % % % % % % % % % % % % % % % % % % % % % % % % % % % % % % % % % % % % % % % % % % % % % % % % % % % % % % % % % % % % % % % % % % % % % % % % % % % % % % % % % % % % % % % % % % 

\section{POVM reconstruction technique for a tree detector}

Alternatively to the classical technique described in the previous section, the POVM of PNRDs can be reconstructed by exploiting strong quantum correlations of twin-beams generated by PDC \cite{D'Ariano2004,Luis99}. In this case, one beam is sent to the photon-number-resolving DUT and the other to a SPAD (with variable quantum efficiency) used as a quantum tomographer \cite{Zam05,zzam05,geno2006,brigen2006,brigen2006bis,brigen2007,moroder2009,brigen2009}.
%, thus yielding a measurement speedup at least $\sqrt{N}$. This alternative retains the statistical reliability advantage of the on-demand Fock state source.
With this technique, significant advantages can be obtained, improving both precision and stability with respect to their classical counterparts.
\par

\subsection{Theory}

First, let us presume that a bipartite system can be prepared in a certain state, described by the density operator $\varrho_R$, and that a known observable with a discrete set of outcomes is measured at the tomographer. As before, our DUT is phase-insensitive, and $\Pi_{nm}$'s are the matrix elements (expressed in the Fock basis) of the POVM to be reconstructed.
\par

The bipartite state in our experiment consists of the optical twin beams $\varrho_R=|R\rangle\rangle\langle\langle R|$, $|R\rangle\rangle = \sum_m R_m |m\rangle|m\rangle$, being $|m\rangle$ the Fock state with $m$ photons and $R_m$ the probability amplitude associated to a particular $|m\rangle$ state.
In this experiment, an "event" is constituted by a detection of $n$ photons at the DUT correlated to the corresponding measurement outcome  of the tomographer (``click" or ``no-click"), which occur with probabilities

\begin{eqnarray}
p(n,{\rm click})=\sum_m \Pi_{nm} |R_m|^2 [1-(1-\eta)^m],  \nonumber\\
\\
p(n,{\rm no-click})=\sum_m \Pi_{nm} |R_m|^2 (1-\eta)^m , \label{probs}\nonumber
\end{eqnarray}
We collect these probabilities for a set of different quantum efficiencies $\eta_{nu}$ ($\nu=$1,...,N), in order to exploit an on/off reconstruction method similar to the one of ref.s \cite{Zam05,zzam05,geno2006,brigen2006,brigen2006bis,brigen2007,moroder2009,brigen2009,Allevi2009,h1,moroder2009,brigen2006bis,Mogil1998}. 

A reliable reconstruction of the $|R_m|^2$ elements can be obtained using the unconditional tomographer no-clik events, which occur with probability $p({\rm no-click}) = \sum_m |R_m|^2 (1-\eta)^m$ \cite{Zam05,zzam05}.
This procedure is simpler than full quantum tomography \cite{Vog89,Dar94,Leo96,Bre97,Lvov2009,Bog2011,Asor2011,dariano03,dariano02}, because it only reconstructs the diagonal elements of the optical state density matrix, thus not needing phase control.
In the following step, the reconstructed $|R_m|^2$ elements are substituted in Eq.s \ref{probs}; the POVM elements can be extracted by inverting the equation system obtained varying the tomographer quantum efficiency. 
\par
\begin{figure}[h]
\includegraphics[height=0.8\columnwidth]{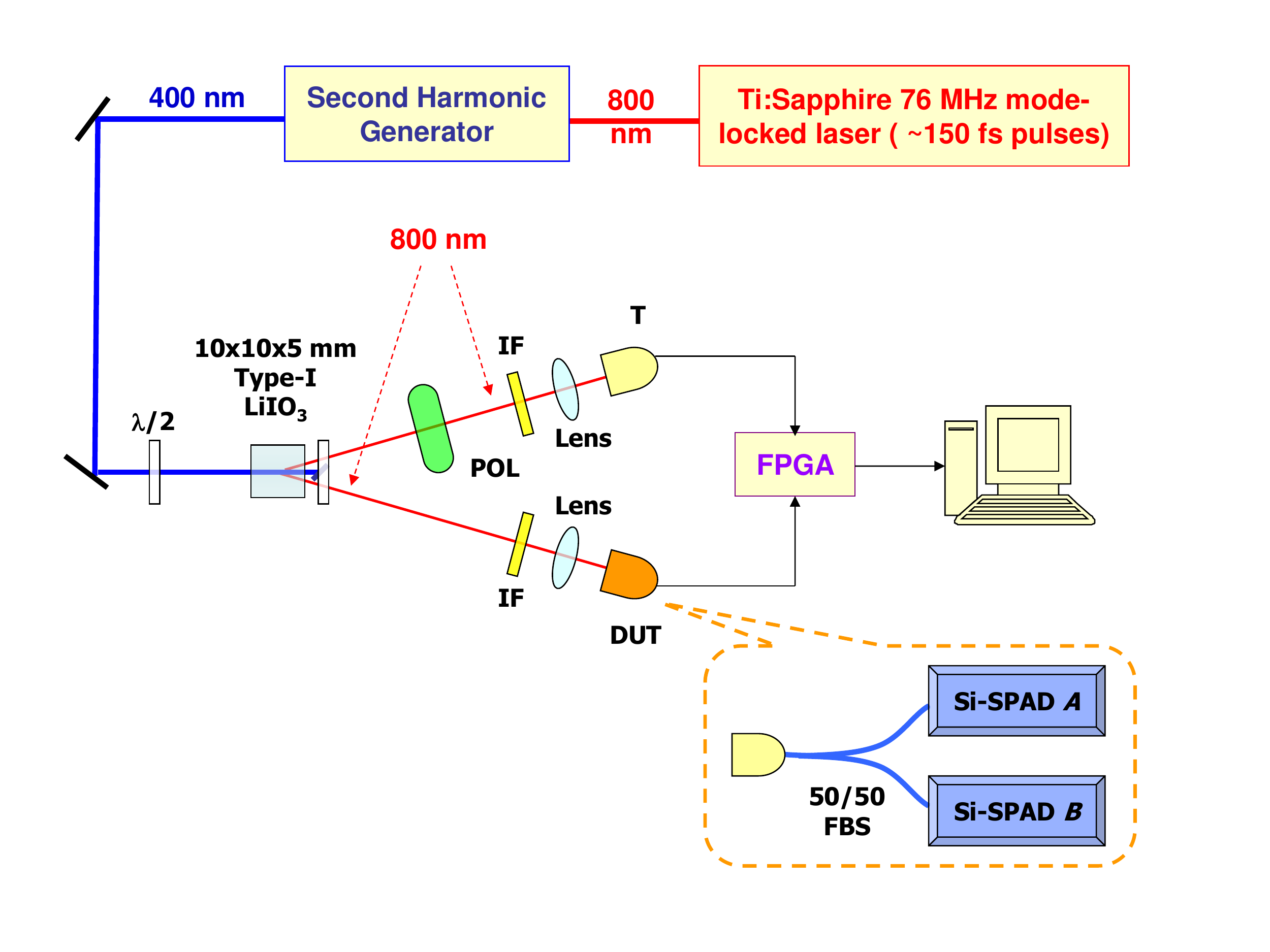}
\vspace{1cm} \caption{Experimental setup: a 800 nm mode-locked laser, doubled via second harmonic generation (SHG) to 400 nm, pumps a LiIO$_3$ crystal producing type-I PDC. One of the generated twin beams is sent to the tomographer (T), while the other is addressed to the DUT. By rotating the linear polarizer on the T-path, the tomographer efficiency is varied. Interference filters (IF) with 20 nm bandwith are used. An FPGA is used for real-time processing and data acquisition.
The DUT (inset) is a detector-tree type PNRD made of two SPADs connected through a 50:50 fibre beam splitter
} \label{f:f1}
\end{figure}
\par

\subsection{Experiment}

The experimental setup \cite{POVM} is shown on Fig. \ref{f:f1}. The twin beam is generated by means of a pulsed Ti-Sapphire laser (76 MHz of repetition rate) at 800 nm. The laser is doubled in frequency and injected into a 10 mm long LiIO$_3$ crystal, producing Type-I PDC. The two beams are addressed respectively to the DUT and the tomographer.
The DUT is a detector tree composed by a 50:50 fiber beam splitter with the outputs connected to two Si-SPADs, thus it can give 3 different outcomes: 0, 1 and 2-or-more detected photons per pulse. Event 0 occurs if neither SPAD clicks, event 1 is registered if either SPAD clicks (but not both) and event 2 corresponds to both SPADs clicking at once.

The two Si-SPADs outputs of the DUT, together with the one of the tomographer  (another Si-SPAD) and a laser trigger pulse, are sent to a Field Programmable Gate Array (FPGA) based data collection and processing system. 
The FPGA is programmed to take data only if the three detectors are available , discarding the events affected by the dead time of the three Si-SPADs.

%%%%%%

\subsection{Results}
The relative frequencies $f(0)$, $f(1)$ and $f(2)$, corresponding to the number of 0-, 1- and 2-click events normalized to their sum, need to be determined to allow the reconstruction of the DUT's POVM.

In addition, for each efficiency $\eta_\nu$ the relative frequencies of conditional events are determined, paired with tomographer's clicks ($f( {\rm click}| 0, \eta_\nu)$, $f( {\rm click}| 1,\eta_\nu)$ $f({\rm click}|2,\eta_\nu)$) and no-clicks ($f({\rm no-click}|0,\eta_\nu)$, $f({\rm no-click}|1,\eta_\nu)$, $f( {\rm no-click}|2,\eta_\nu)$). As mentioned before, the reconstruction of the photon number distribution of the bipartite state is the first step in obtaining the POVM elements: the $|Rm|^2$ elements are extracted exploiting the no-click frequencies of the tomographer, $f(no-click, \eta_{\nu})=\sum{^2_{i=0}}f(no-click|i, \eta_{\nu})$
\begin{figure}
\includegraphics[width=0.8\columnwidth]{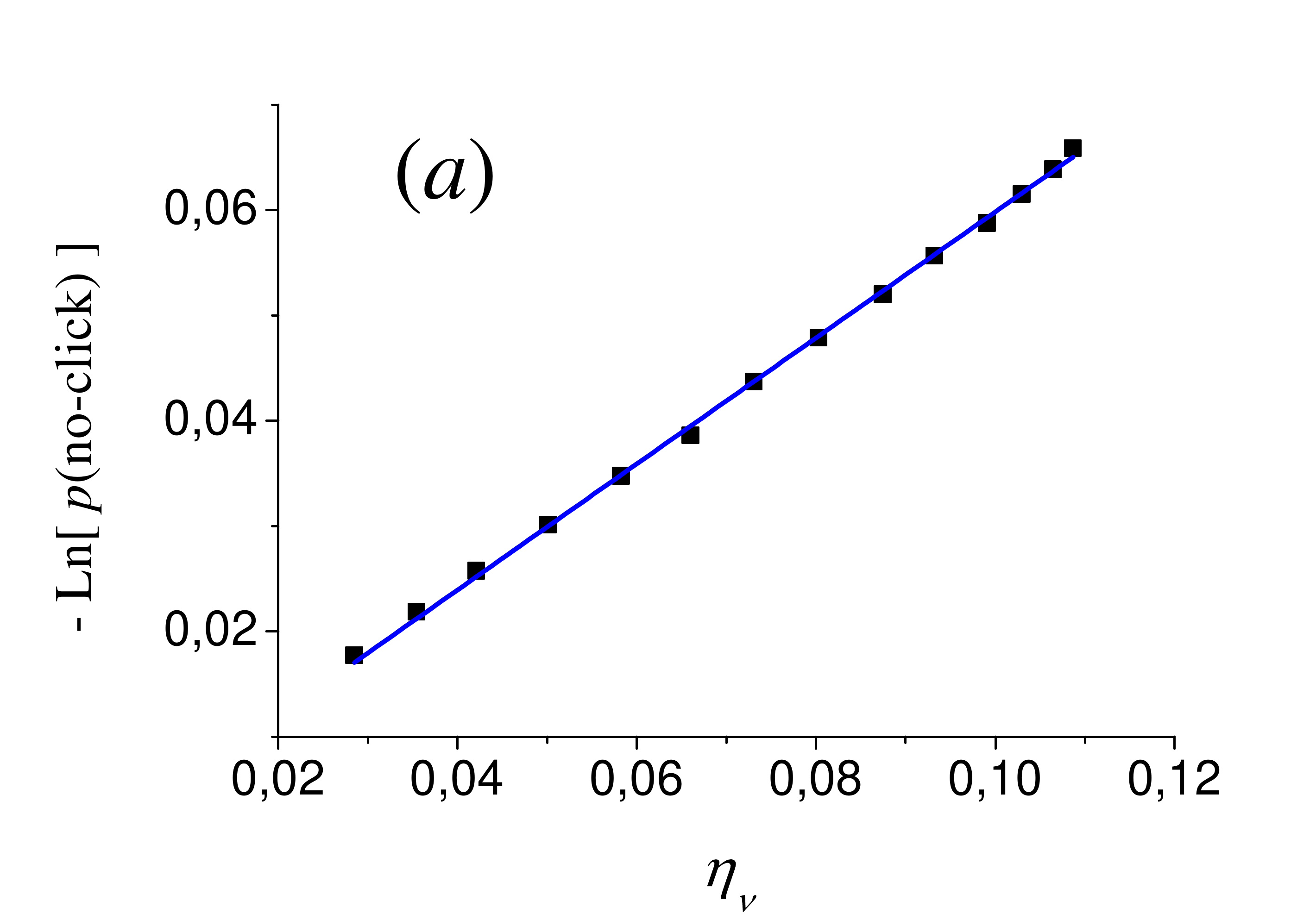}\\
\includegraphics[width=0.8\columnwidth]{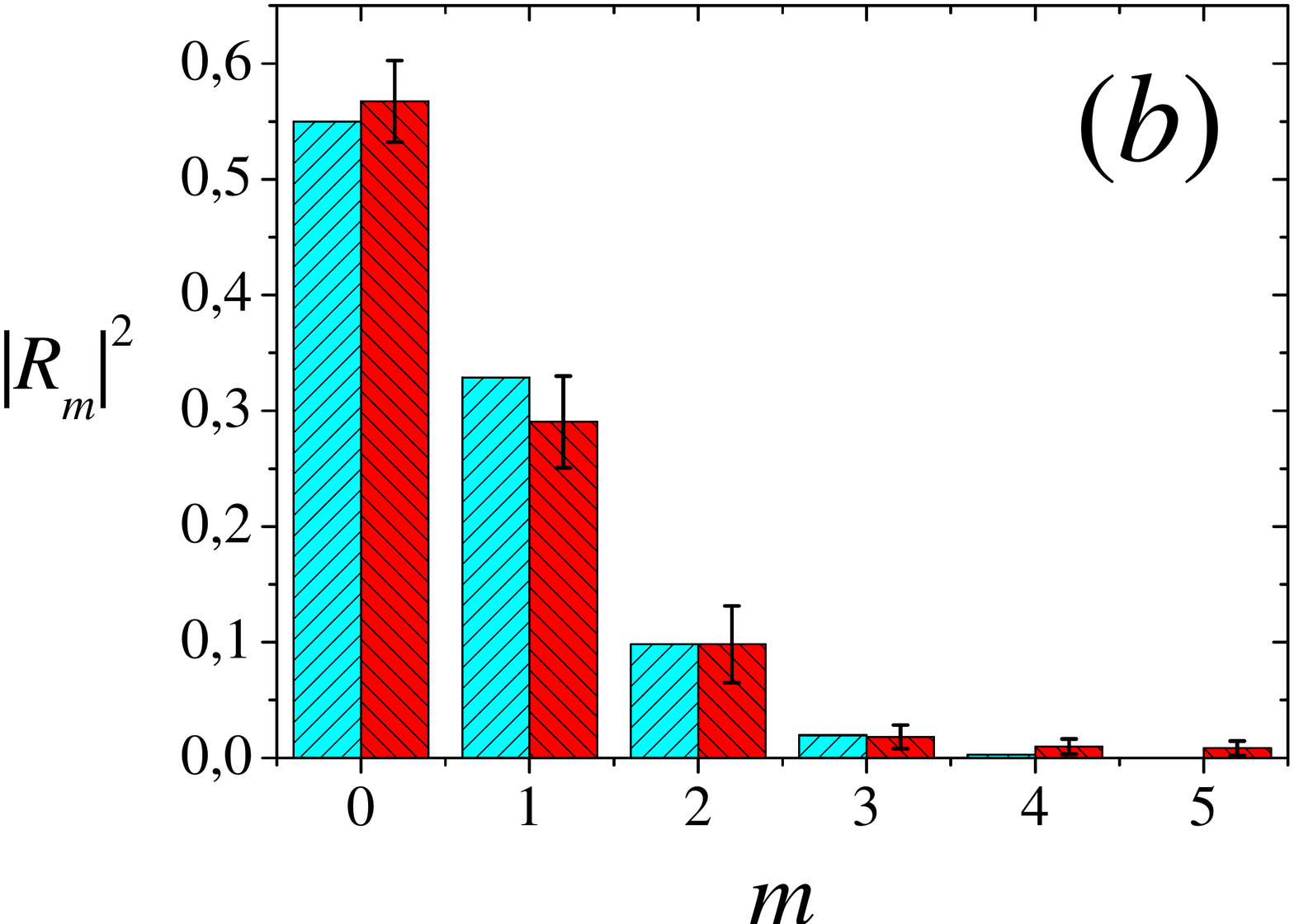}
\caption{(a) The blue line is the best fit of the experimental $p({\rm no-click})$ (black dots) given by a Poisson distribution with $\mu=0.5983 \pm 0.0017$ average photons per pulse. (b) Comparison between the reconstructed distribution (red bars), with a Poisson distribution (light blue bars) with the photon number determined by the fit in (a). The uncertainties represent the $1\sigma$ variations in the reconstructions performed on 30 different data sets. Since in this experiment the probability of observing 5 or more photons is negligible (less than 4 $\times 10^{-4}$), data are shown only up to $m=5$ photons.}
   \label{stato}
   \end{figure}
\par
Fig. \ref{stato} shows that, as expected, the experimentally reconstructed photon distribution is
in excellent agreement with the Poisson distribution, with a fidelity larger than $99.4\%$.

By substituting the $|R_m|^2$'s together with the set of calibrated efficiencies $\{\eta_\nu \}$ into Eq. (\ref{probs}), the quantities $\Pi_{nm}$ are reconstructed using a regularized least square method \cite{lun09, tomo} to minimize the deviation between the measured and theoretical values of the probabilities. 

In particular, for each $\eta_{\nu}$ and for each output $n$ of the DUT, the deviation between the observed $p_{\mathrm{exp}}(n,{\rm click}) = f(n) f({\rm click}|n, \eta_\nu)$ and theoretical probabilities $p(n,{\rm click})$ is minimized, as well as the deviation between $p_{\mathrm{exp}}(n,{\rm no-click}) = f(n) f({\rm no-click}|n, \eta_\nu)$ and $p(n,{\rm no-click})$.

\begin{figure}
     \begin{center}
        \subfigure{%
            \includegraphics[width=0.4\textwidth]{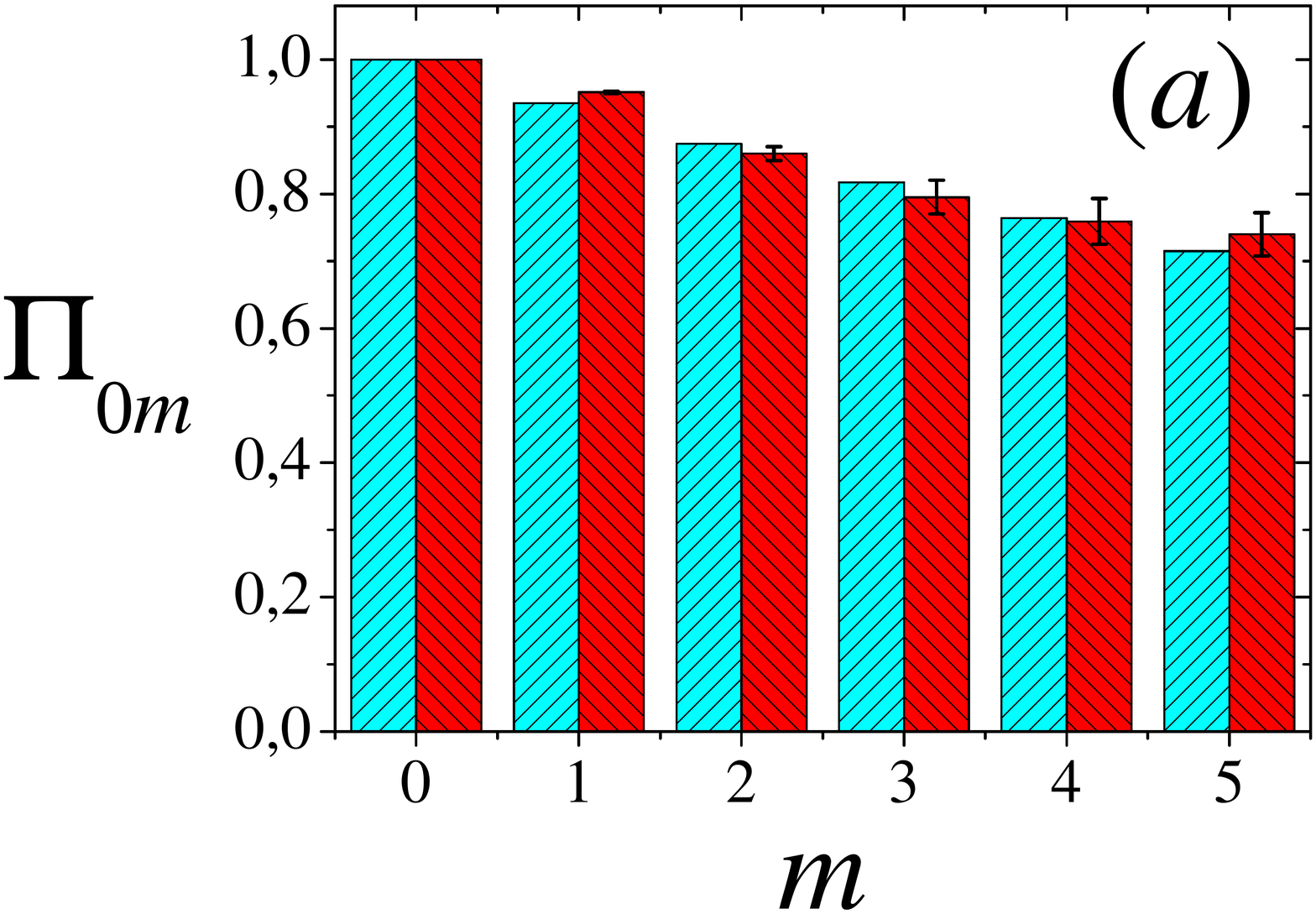}
        }\\
        \subfigure{%
           \includegraphics[width=0.4\textwidth]{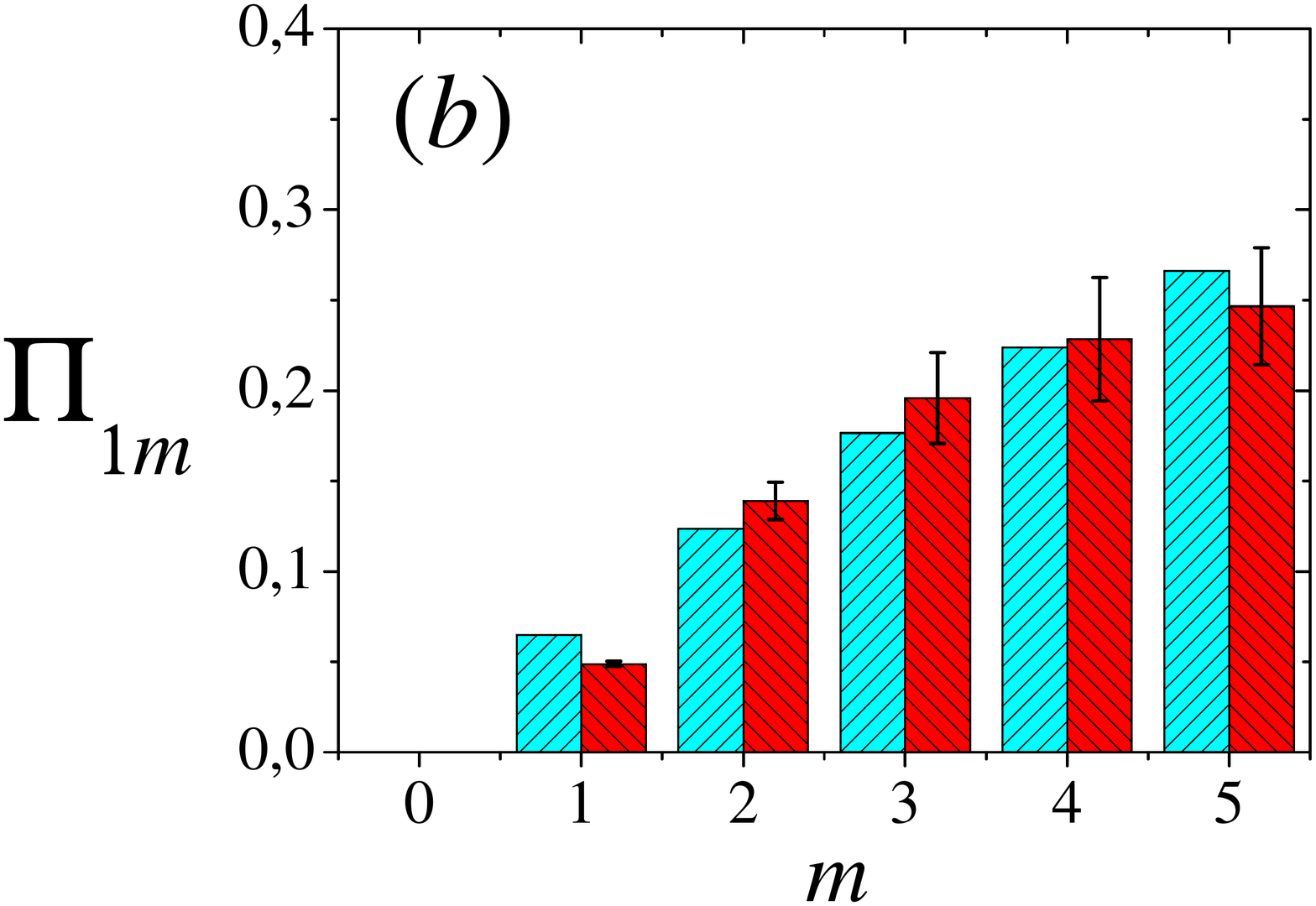}
        }    
        \subfigure{%
           \includegraphics[width=0.4\textwidth]{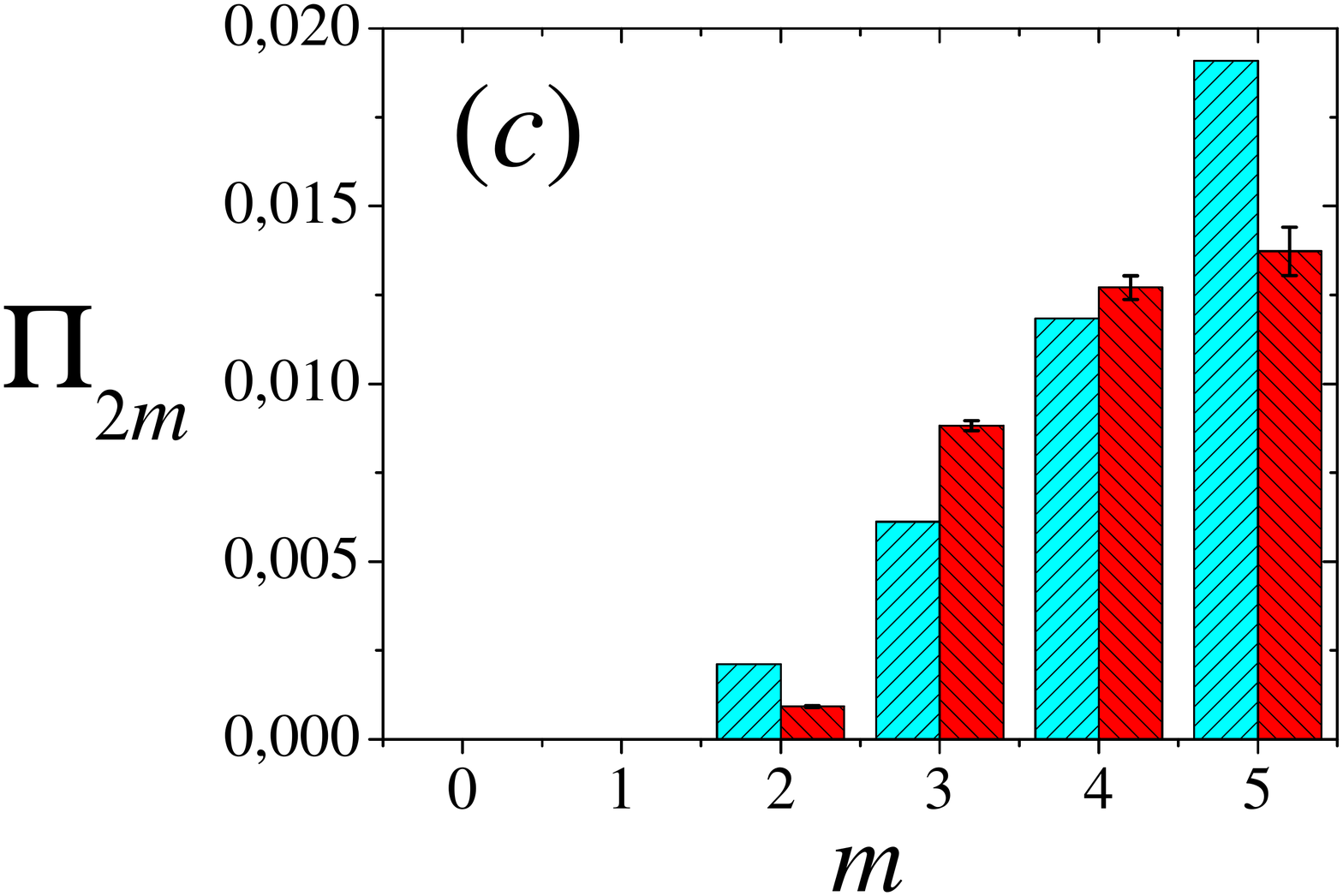}
                }  
    \end{center}
    \caption{POVM elements reconstructed up to $m=5$.
    Experimental reconstructed histograms are shown in red, theoretical histograms are shown in light blue for: (a) $\Pi_{0m}$, (b) $\Pi_{1m}$ and (c) $\Pi_{2m}$. Quality of reconstruction of POVM elements with $m<5$ is independently confirmed by observed fidelities above $99.9 \%$. The accuracy starts deteriorating for input states with $m\geq5$. Statistical fluctuations in the reconstructions performed on 30 different data sets are represented by the uncertainty bars.
     }
   \label{povm}
\end{figure}
The reconstructed $\Pi_{0m}$, $\Pi_{1m}$, $\Pi_{2m}$ are presented in Fig. \ref{povm}. 
Excellent agreement between theoretical and experimental results are supported by the high fidelities (above 99.9 \%) for values  $m < 5$, while for $m\geq5$ the quality of the POVM reconstruction rapidly decreases. 
\par

\begin{figure}
\includegraphics[width=0.95\columnwidth]{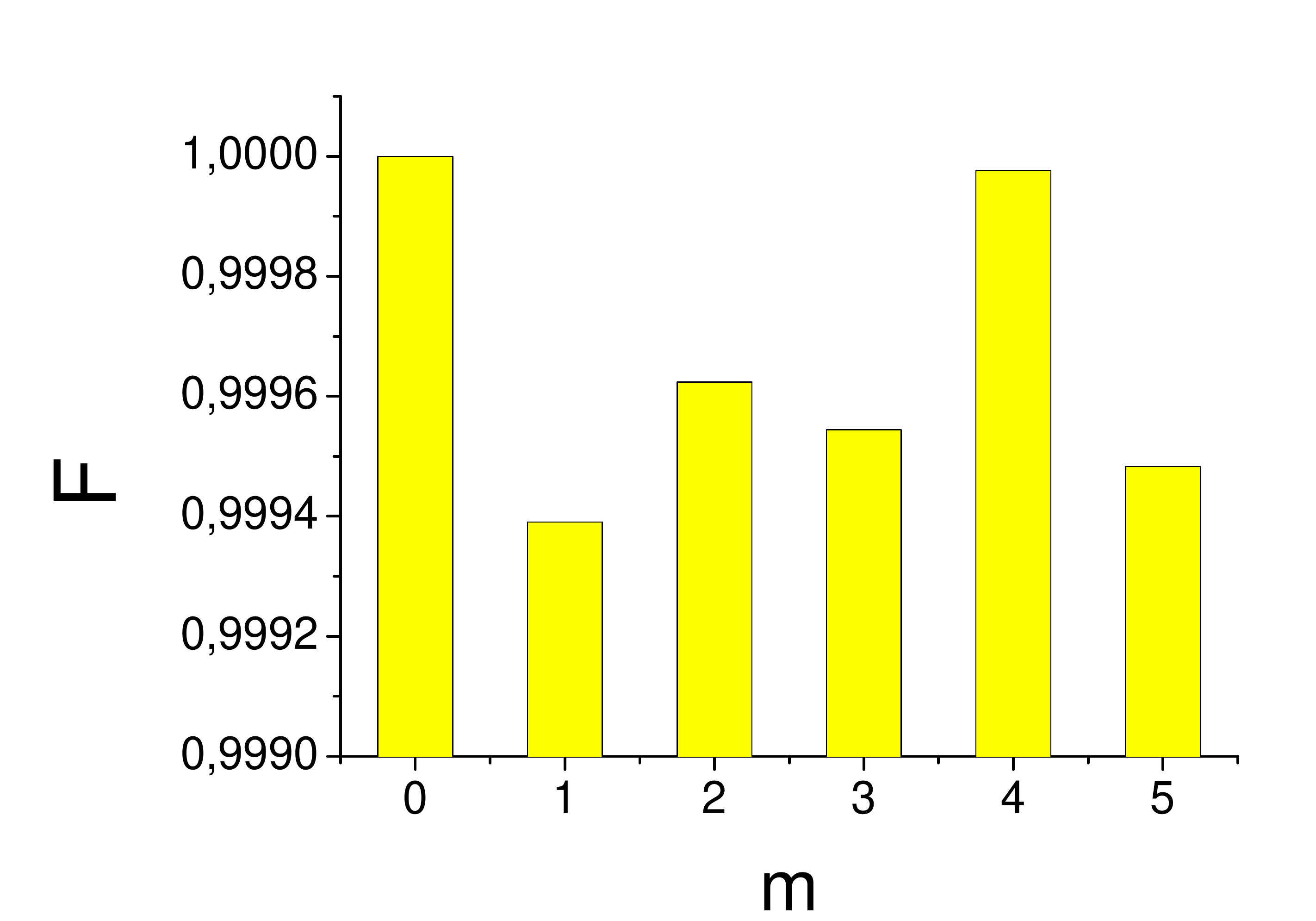}
\caption{Fidelities of the reconstructed POVM entries (with respect to the theoretical model) for each $m$. The reliability of the reconstruction is confirmed by the fact that the fidelities are all above $99.9\%$.} \label{experiment}
\end{figure}

In Fig. \ref{experiment} are reported the fidelities of the reconstructed POVM elements shown in Fig. \ref{povm}: the high values obtained confirm that the extracted POVM provides a reliable quantum description of the detection process.
%%%%%

\section{Conclusions}

%In this paper we have reviewed some recent progress about calibration of detectors performed at INRIM. We hope that this review can help the reader in achieving a first idea about the problems and perspectives of this kind of studies, fostering further researches in the field, that represents a significant part of present studies in photonic quantum technologies.
In this review, we have presented some recent progresses achieved in the INRIM Quantum Optics labs about calibration of single or few photon detectors. We hope that the illustrated results can give the reader an idea about the problems, the state of the art and the perspectives of this kind of studies, fostering further efforts in these directions. We also stress that, in our opinion, such a topic is of crucial relevance for the present and future developments in photonic quantum technologies and related research fields (e.g. quantum information and cryptography \cite{QMAVELLA}).

The research leading to these results has received funding from the European Union on the basis of Decision No. 912/2009/EC (project IND06-MIQC), from MIUR (FIRB Grants No. RBFR10UAUV, No. RBFR10VZUG and No. RBFR10YQ3H, and Progetto Premiale P5 "Oltre i limiti classici della misura"), and from Compagnia di San Paolo.

%\bibliographystyle{unsrt}

%\bibliography{bibliografiauno}

\begin{thebibliography}{99}

\bibitem{Genovese2005} M. Genovese, \emph{Phys. Rep.} 413, 319 {\bf (2005)}
\bibitem{Croca2013} J. R. Croca , \emph{Quantum Matter} 2, 1 {\bf (2013)}
\bibitem{QMAVELLA}  P. Traina, M.Gramegna,  A. Avella, A. Cavanna, D. Carpentras, I. P. Degiovanni, G. Brida and M. Genovese, \emph{Quantum Matter} 2, 153 {\bf (2013)}
\bibitem{Scarani2009} V. Scarani, H. Bechmann-Pasquinucci, N. J. Cerf, M. Dusek, N. Lutkenhaus and M. Peev,  \emph{Rev. Mod. Phys.} 81, 1301 {\bf (2009)}
\bibitem{Gisin2002} N. Gisin, G. Ribordy, W. Tittel and H. Zbinden,  \emph{Rev. Mod. Phys.} 74, 145 {\bf (2002)}
\bibitem{Braunstein2005}  S. L. Braunstein and P. van Loock, \emph{Rev. Mod. Phys.} 77, 513 {\bf (2005)}
\bibitem{Ladd2010}  T. D. Ladd, F. Jelezko, R. Laflamme, Y. Nakamura, C. Monroe and J. L. O'Brien, \emph{Nature} 464, 45 {\bf (2010)}
\bibitem{Kok2007} P. Kok, W. J. Munro, K. Nemoto, T. C. Ralph, J. P. Dowling and G. J. Milburn,  \emph{Rev. Mod. Phys.} 79, 135 {\bf (2007)}
\bibitem{Zwinkels10} J. C. Zwinkels, E. Ikonen, N. P. Fox, G. Ulm, M. L. Rastello, \emph{Metrologia} 47, R15 {\bf (2010)}
\bibitem{Giovannetti2011} V. Giovannetti, S. Lloyd and L. Maccone,  \emph{Nat. Phot.} 5, 222 {\bf (2011)}


\bibitem{VV1} J. Perina, O. Haderka, V. Michalek and M. Hamar, \emph{Optics Letters} 37, 2475 {\bf (2012)}
\bibitem{VV2} D. A. Kalashnikov, Si-Hui Tan and L. A. Krivitsky, \emph{Optics Express} 20, 5044 {\bf (2012)}
\bibitem{josab} G. Brida, M. Genovese, I. Ruo-Berchera, M. Chekhova and A. Penin \emph{JOSA B} 23, 2185 {\bf (2006)}
\bibitem{JMO2009} G. Brida, A. Meda, M. Genovese, E. Predazzi and I. Ruo-Berchera, \emph{Journal of Modern Optics} 56, 201 {\bf (2009)}
\bibitem{optex2008} G. Brida, M. Chekhova, M. Genovese and I. Ruo-Berchera, \emph{Optics Express} 16, 12550 {\bf (2008)}
\bibitem{Lindenthal} M. Lindenthal and J. Kofler \emph{Applied Optics} 45, 6059 {\bf (2006)}
\bibitem{Agafonov} I. N. Agafonov, M.V.Chekhova, T. S. Iskhakov, A. N. Penin, G. O. Rytikov and O. A. Shumilkina,  \emph{IJQI} 9, 251 {\bf (2011)}


\bibitem{burle} G. Zambra, M. Bondani, A. S. Spinelli and  A. Andreoni, \emph{Rev. Sci. Instrum.} 75, 2762 {\bf (2004)}
\bibitem{Allevi2010} A. Allevi, M. Bondani, and A. Andreoni  \emph{Opt. Lett.} 35, 1707 {\bf (2010)}
\bibitem{zhang2012} G. Q. Zhang, X. J. Zhai, C. J. Zhu, H. C. Liu and Y. T. Zhang, \emph{IJQI} 10, 1230002 {\bf (2012)}
\bibitem{NIST} M. Ramilli, A. Allevi, A. Chmill, M. Bondani,  M. Caccia and A. Andreoni,  \emph{J. Opt. Soc. Am. B} 27, 852 {\bf (2010)}
\bibitem{Haderka2005} O. Haderka, J. Peřina Jr., M. Hamar, and J. Peřina,  \emph{\PRA} 71, 033815 {\bf (2005)}
\bibitem{Kalash2011} D. A. Kalashnikov, S. H. Tan, M. V. Chekhova and L. A. Krivitsky,  \emph{Opt. Exp.} 19, 9352 {\bf (2011)}
\bibitem{lapo2011} L. Lolli, G. Brida, I. P. Degiovanni, M. Gramegna, E. Monticone, F. Piacentini, C. Portesi, M. Rajteri, I. Ruo Berchera, E. Taralli and P. Traina, \emph{ Int. J. Quantum Inform.} 9, 405 {\bf (2011)}
\bibitem{fuk11} D. Fukuda, G. Fujii, T. Numata, K. Amemiyaand,  A. Yoshizawa, H. Tsuchida, H. Fujino, H. Ishii, T. Itatani,  S. Inoue and T. Zama,  \emph{Opt. Express} 19, 870 {\bf (2011)}
\bibitem{pre09} D. Prele, M. R. Piat, E. L. Breelleand, F. Voisin,  M. Pairat, Y. Atik, B. Belier,  L. Dumoulin, C. Evesque, G. Klisnick,  S. Marnieros, F. Pajot, M. Redonand and G. Sou, \emph{IEEE T. Appl. Supercon.} 19, 501 {\bf (2009)}
\bibitem{cab08} B. Cabrera \emph{J. Low Temp. Phys.} 151, 82 {\bf (2008)}
\bibitem{lit08}  A. E. Lita, A. J. Miller and S. W. Nam, \emph{Opt. Express} 16, 3032 {\bf (2008)}
\bibitem{ros05} D. Rosenberg, A. E. Lita, A. J. Millerand and  S. W. Nam, \emph{\PRA} 71, 061803 {\bf (2005)}
\bibitem{bandler06} S. R. Bandler, E. Figueroa-Feliciano, N. Iyomoto, R. L. Kelley, C. A. Kilbourne, K. D. Murphy,  F. S. Porter, T. Saab,  and J. Sadleir, \emph{Nucl. Instrum. Meth. A} 559, 817 {\bf (2006)}
\bibitem{multiTemporal} D. Achilles, Ch. Silberhorn, C. Sliwa, K. Banaszek and I. A. Walmsley \emph{Opt. Lett.} 28, 2387 {\bf (2003)}



\bibitem{fitch2003}  M. J. Fitch, B. C. Jacobs, T. B. Pittman, and J. D. Franson, \emph{\PRA} 68, 043814 {\bf (2003)}
\bibitem{rehacek2003} J. Rehacek, Z. Hradil, O. Haderka, J. Perina, Jr., and M. Hamar, \emph{\PRA} 67, 061801 {\bf (2003)}
\bibitem{hederka2004} O. Haderka, M. Hamar, J. Perina Jr \emph{EPJD} 28, 149 {\bf (2004)}
\bibitem{multiSpatial} L. A. Jiang, E. A. Dauler and J. T. Chang, \emph{Phys. Rev. A} 75, 062325 {\bf (2007)}
\bibitem{multiSpatial2} A. Divochiy, F. Marsili, D. Bitauld, A. Gaggero, R. Leoni, F. Mattioli, A. Korneev, V. Seleznev,  N. Kaurova, O. Minaeva, G. Gol'tsman, K G. Lagoudakis, M. Benkhaoul, F. Le\`vy and   A. Fiore,  \emph{Nat. Photonics} 2, 302 {\bf (2008)}
\bibitem{Burnham1970}  D. C. Burnham and D. L. Weinberg, \emph{Phys. Rev. Lett.} 25, 84 {\bf (1970)}
\bibitem{Klyshko1977} D. N. Klyshko \emph{Sov. J. Quantum Electron.} 7, 591 {\bf (1977)}
\bibitem{Kwiat1994}  P. G. Kwiat, A. M. Steinberg, R. Y. Chiao, P. H. Eberhard and M. D. Petroff, \emph{Appl. Opt.} 33, 1844 {\bf (1994)}
\bibitem{Migdall1995} A.L. Migdall, R.U. Datla, A. Sergienko and Y.H. Shih, \emph{Metrologia} 32, 479 {\bf (1996)}
\bibitem{Dauler1998} E. Dauler, A. Migdall, N. Boeuf, R. Datla, A. Muller and A. Sergienko, \emph{Metrologia} 35, 259 {\bf (1998)}
\bibitem{Brida2000} G. Brida, S. Castelletto, I. P. Degiovanni, C. Novero and M. L. Rastello, \emph{Metrologia} 37, 625 {\bf (2000)}
\bibitem{Castelletto2002} S. Castelletto, I. P. Degiovanni and M. L. Rastello, \emph{J. Opt. Soc. Am. B} 19, 1247 {\bf (2002)}
\bibitem{Ghazi2000} A. Ghazi-Bellouati, A. Razet, J. Bastie, M. E. Himbert, I. P. Degiovanni, S. Castelletto and M. L. Rastello,  \emph{Metrologia} 42, 271 {\bf (2005)}
\bibitem{Migdall2002} A. Migdall, S. Castelletto, I. P. Degiovanni and M. L. Rastello, \emph{Appl. Opt.} 41, 2914 {\bf (2002)}
\bibitem{Polyakov2007} S. V. Polyakov and A.L. Migdall \emph{Optics Express} 15, 1390 {\bf (2007)}
\bibitem{jessica} J. Y. Cheung, C. J. Chunnilall, G. Porrovecchio, M. Smid and E. Theocharous, \emph{Opt. Express} 19, 20347 {\bf (2011)}
\bibitem{penin1991} A. N. Penin and A. V. Sergienko, \emph{Appl. Opt.} 30, 3582 {\bf (1991)}
\bibitem{brigrn2000} G. Brida, M. Genovese and C. Novero, \emph{Journal of Modern Optics} 47, 2099 {\bf (2000)}
\bibitem{gram2006} G. Brida, M. Genovese and M. Gramegna,  \emph{Laser Phys. Lett.} 3, 115 {\bf (2006)}
\bibitem{malygin} A. A. Malygin, A. N. Penin and A. V. Sergienko,  \emph{Sov. J. Quantum Electron.} 11, 939 {\bf (1981)}
\bibitem{Kly1980} D. N. Klyshko, \emph{Sov. J. Quantum Electron.} 10, 1112 {\bf (1980)}
\bibitem{Bowman1986} S. R. Bowman, Y. H. Shih and C. O. Alley, \emph{Proc. SPIE} 633, 24 {\bf (1986)}
\bibitem{Ginzburg1993} V. M. Ginzburg, N. G. Keratishvili, Ye. L. Korzhenevich , G. V. Lunev and A. N. Penin, \emph{Metrologia} 30, 367 {\bf (1993)}
\bibitem{cheung2004} J. Y. Cheung, M. P. Vaughan, J. R. Mountford and C. J. Chunnilall, \emph{Proc. SPIE} 5161, 365 {\bf (2004)}

\bibitem{Luisell1961} W. H. Louisell and A. Yariv, \emph{Phys. Rev.} 124, 1646 {\bf (1961)}
\bibitem{Rarity1987} J. G. Rarity, K. D. Ridley and P. R. Tapster, \emph{ Appl. Opt.} 26, 4616 {\bf (1987)}
\bibitem{brigenovese2005} G. Brida, M. Genovese, M. Gramegna, M. L. Rastello, M. Chekhova, and L. Krivitsky \emph{JOSA B} 22, 488 {\bf (2005)}



\bibitem{brideg2008} G. Brida, I. P. Degiovanni, M. Genovese, V. Schettini, S. V. Polyakov, and A. Migdall, \emph{Opt. Exp.} 16, 11750 {\bf (2008)}
\bibitem{8} D.N. Klyshko, \emph{Photons and Nonlinear Optics}, Gordon and Breach Science Publishers {\bf (1988)} 

\bibitem{10} G. Brida, S. Castelletto, C. Novero and M. L. Rastello, \emph{Metrologia} 35, 397 {\bf (1998)}
\bibitem{11} A. Migdall, \emph{Physics Today} 52, 41 {\bf (1999)}
\bibitem{16} Joint Committee for Guides in Metrology  \emph{Evaluation of measurement data — Guide to the expression of uncertainty in measurement}, BIPM,  {\bf (2008)}
\bibitem{mingolla} M. G. Mingolla, G.Brida, paper in preparation 
\bibitem{castelletto2000} S. Castelletto, I. P. Degiovanni and M. L. Rastello,  \emph{Metrologia} 37, 613 {\bf (2000)}
\bibitem{Avella2011} A. Avella, G. Brida, I. P. Degiovanni, M. Genovese, M. Gramegna, L. Lolli, E. Monticone, C. Portesi, M. Rajteri, M. L. Rastello, E. Taralli, P. Traina and M. White, \emph{Opt. Express} 19, 23249 {\bf (2011)}

\bibitem{fab} C. Portesi, E. Taralli, R. Rocci, M. Rajteri and E. Monticone, \emph{J. Low Temp. Phys} 151, 261 {\bf (2008)}
\bibitem{taralli} C. Portesi, L. Lolli, E. Monticone, M. Rajteri, I. Novikov and J. Beyer, \emph{Supercond. Sci. Technol.} 23, 105012 {\bf (2010)}
\bibitem{etf} K. D.  Irwin \emph{Appl. Phys. Lett.} 66, 1998 {\bf (1995)}
\bibitem{lapo} L. Lolli, E. Taralli, C. Portesi, D. Alberto, M. Rajteri  and E. Monticone, \emph{IEEE Trans. Appl. Supercond.} 21, 215 {\bf (2011)}
\bibitem{squid}  D. Drung, C. Assmann, J. Beyer, A. Kirste, M. Peters, F. Ruede and T. Schurig, \emph{IEEE Trans. Appl. Supercond.} 17, 699 {\bf (2007)}
\bibitem{qurad2} S. V. Polyakov and A. L. Migdall, \emph{J. Mod. Opt} 56, 1045 {\bf (2009)}
\bibitem{tomo} G. Brida, L. Ciavarella, I. P. Degiovanni, M. Genovese, L. Lolli, M. G. Mingolla, F. Piacentini, M. Rajteri, E. Taralli and M. G. A. Paris, \emph{New J. Phys.} 14, 085001 {\bf (2012)}
\bibitem{had} R. H. Hadfield, \emph{Nature Photon.} 3, 636 {\bf (2009)}
\bibitem{had1} C. Silberhorn, \emph{Contemp. Phys.} 48, 143 {\bf (2007)}
\bibitem{lss99}  A. Luis and L. L. Sanchez-Soto, \emph{Phys. Rev. Lett.} 83, 3573 {\bf (1999)}
\bibitem{jar01} J. Fiurasek,  \emph{Phys. Rev. A} 64, 024102 {\bf (2001)}
\bibitem{dem03}  F. Demartini, A. Mazzei, G. M. Ricci, and G. M. D'Ariano, \emph{\PRA} 67, 062307 {\bf (2003)} 
\bibitem{mit03}   M. W. Mitchell, C. W. Ellenor, S. Schneider and A. M. Steinberg, \emph{Phys. Rev. Lett.} 91, 120402 {\bf (2003)} 
\bibitem{Zam05} A. R. Rossi, S. Olivares  and M. G. A. Paris, \emph{Phys. Rev. A} 70, 055801 {\bf (2004)}
\bibitem{zzam05} G. Zambra, M. Bondani, A. Andreoni, M. Gramegna, M. Genovese, G. Brida, A. Rossi and M. G. A. Paris, \emph{Phys. Rev. Lett.} 95, 063602 {\bf (2005)}
\bibitem{lob08} M. Lobino, D. Korystov, C. Kupchak, E. Figueroa, B. C. Sanders and A. I. Lvovsky,  \emph{Science} 322, 563 {\bf (2008)}
\bibitem{rah11} S. Rahimi-Keshari, A. Scherer, A. Mann, A. T. Rezakhani, A. I. Lvovsky, B. C. Sanders, \emph{New J. Phys.} 13, 013006 {\bf (2011)}
\bibitem{D'Ariano2004} G. M. D'Ariano, L. Maccone and P. Lo Presti \emph{Phys. Rev. Lett.} 93, 250407 {\bf (2004)}
\bibitem{h1}  Z. Hradil, D. Mogilevtsev and J. Rehacek,  \emph{Phys. Rev. Lett.} 96, 230401 {\bf (2006)}
\bibitem{lun09} J. S. Lundeen, A. Feito,  H. Coldenstrodt-Ronge,  K. L. Pregnell,  Ch. Silberhorn, T. C. Ralph, J. Eisert,  M. B. Plenio and I. A.Walmsley, \emph{Nat. Phys.}5 , 27 {\bf (2009)}
\bibitem{h2} J. Rehacek, D. Mogilevtsev and Z. Hradil,  \emph{Phys. Rev. Lett.} 105, 010402 {\bf (2010)}
\bibitem{por08} C. Portesi, E. Taralli, R. Rocci , M. Rajteri  and
E. Monticone,  \emph{J. Low Temp. Phys.} 151, 261 {\bf (2008)}
\bibitem{tar07} E. Taralli, M. Rajteri, E. Monticone and C. Portesi, \emph{Int. J. Quantum. Inf.} 5, 293 {\bf (2007)}
\bibitem{dru07} D. Drung, C. Assmann, J. Beyer, A. Kirste, M. Peters, F. Ruede  and Th. Schurig,  \emph{IEEE T. Appl. Supercon.}  17, 699 {\bf (2007)}

\bibitem{POVMTES} G. Brida, L. Ciavarella, I. P. Degiovanni, M. Genovese, L. Lolli, M. G. Mingolla, F. Piacentini, M. Rajteri, E. Taralli and M. G. A. Paris,  \emph{New J. Phys.} 14, 085001 {\bf (2012)}
\bibitem{Luis99} A. Luis and L.L. Sanchez-Soto, \emph{Phys. Rev. Lett.} 83, 18 {\bf (1999)}
\bibitem{geno2006} M. Genovese, G. Brida, G. Zambra,  A. Andreoni, M. Bondani, M. Gramegna, A. Rossi and M. G. A. Paris, \emph{ Laser Physics} 16, 385 {\bf (2006)}
\bibitem{brigen2006} G. Brida, M. Genovese,  M. Gramegna, M. G. A. Paris, E. Predazzi and E. Cagliero, \emph{Open Syst. \& Inf. Dyn.} 13, 333 {\bf (2006)}
\bibitem{brigen2006bis}G. Brida, M. Genovese, M.G.A. Paris and F. Piacentini,  \emph{ Opt. Lett.} 31, Issue 23, 3508 {\bf (2006)}
\bibitem{brigen2007} G. Brida, M. Genovese, M. G. A. Paris, F. Piacentini, E. Predazzi and E. Vallauri,  \emph{Opt. \& Spect. } 103, 95 {\bf (2007)}
\bibitem{moroder2009} T. Moroder, M. Curty and N. Lütkenhaus,  \emph{New J. Physics} 11, 045008  {\bf (2009)}
\bibitem{brigen2009} G. Brida, M. Genovese, A. Meda, S. Olivares, M. G. A. Paris and F. Piacentini, \emph{Journ. Mod. Opt.} 56, 196 {\bf (2009)}

\bibitem{Allevi2009} A. Allevi, A. Andreoni, M. Bondani, G. Brida, M. Genovese, M. Gramegna, P. Traina, S. Olivares, M. G. A. Paris and G. Zambra, \emph{\PRA} 80, 022114 {\bf (2009)}
\bibitem{Mogil1998} D. Mogilevtsev, Z. Hradil and J. Perina, \emph{Quantum. Semicl. Opt.} 10, 345 {\bf (1998)}



\bibitem{Vog89} K. Vogel and H. Risken,  \emph{Phys Rev. A} 40, 2847 {\bf (1989)}
\bibitem{Dar94} G. D'Ariano, C. Macchiavello and M. G. A. Paris, \emph{Phys Rev. A} 50, 4298 {\bf (1994)}
\bibitem{Leo96}  U. Leonhardt, M. Munroe, T. Kiss, T. Richter and M. G. Raymer \emph{Opt. Commun.} 127, 144 {\bf (1996)}


\bibitem{Bre97} G. Breitenbach, S. Schiller and J. Mlynek, \emph{Nature} 387 , 471 {\bf (1997)}

\bibitem{Lvov2009} A. I. Lvovsky, M. G. Raymer, \emph{ Rev. Mod. Phys.} 81, 299 {\bf (2009)}
\bibitem{Bog2011} Yu. I. Bogdanov, G. Brida, I. D. Bukeev, M. Genovese, K. S. Kravtsov, S. P. Kulik, E. V. Moreva, A. A. Soloviev and A. P. Shurupov, \emph{\PRA} 84, 042108 {\bf (2011)}
\bibitem{Asor2011} M. Asorey, P. Facchi, G. Florio, V.I. Man'ko, G. Marmo, S. Pascazio and E.C.G. Sudarshan, \emph{Phys. Lett. A} 375, 861 {\bf (2011)}
\bibitem{dariano03} G. M. D'Ariano, M. G. A. Paris, M. F. Sacchi, \emph{Adv. in Im. and El. Phys.}128, 205 {\bf (2003)}
\bibitem{dariano02} G. M. D'Ariano, M. De Laurentis, M. G. A. Paris, A. Porzio and S. Solimeno,\emph{J. Opt. B} 4, 127 {\bf (2002)}
\bibitem{POVM} G. Brida, L. Ciavarella, I. P. Degiovanni, M. Genovese, A. Migdall, M. G. Mingolla, M. G. A. Paris, F. Piacentini, and S. V. Polyakov,   \emph{\PRL} 108, 253601 {\bf (2012)}


\end{thebibliography}

\end{document}